\documentclass[prl,aps,twocolumn,floatfix,showpacs,10pt]{revtex4-1}
\usepackage[colorlinks=true,linkcolor=blue,anchorcolor=red,citecolor=blue,urlcolor=blue]{hyperref}
\usepackage{amsmath}
\usepackage{pdfpages}
\usepackage{graphicx}
\usepackage{pdfpages}
\usepackage{bm}
\usepackage{subfigure}

\newcommand{\bk}{{\bf k}}

\newcommand{\bn}{{\bf n}}

\newcommand{\bp}{{\bf p}}

\newcommand{\bv}{{\bf v}}
\newcommand{\bq}{{\bf q}}
\newcommand{\bA}{{\bf A}}

\newcommand{\ket}[1]{\left| #1 \right>} 
\newcommand{\bra}[1]{\left< #1 \right|} 

\begin{document}
\title{Detecting monopole charge in Weyl semimetals via quantum interference transport}
\author{Xin Dai$^1$, Hai-Zhou Lu$^{2,*}$, Shun-Qing Shen$^{3}$, and Hong Yao$^{1,4,}$}
\email{Corresponding authors: luhz@sustc.edu.cn}
\email{yaohong@tsinghua.edu.cn}
\affiliation{$^1$Institute for Advanced Study, Tsinghua University, Beijing 100084, China\\
$^2$Department of Physics, South University of Science and Technology of China, Shenzhen 518055, China\\
$^3$Department of Physics, The University of Hong Kong, Pokfulam Road, Hong Kong, China\\
$^4$Collaborative Innovation Center of Quantum Matter, Beijing 100084, China}

\date{\today}
\begin{abstract}
Topological Weyl semimetals can host Weyl nodes with monopole charges in momentum space. How to detect the signature of the monopole charges in quantum transport remains a challenging topic. Here, we reveal the connection between the parity of monopole charge in topological semimetals and the quantum interference corrections to the conductivity. We show that the parity of monopole charge determines the sign of the quantum interference correction, with odd and even parity yielding the weak antilocalization and weak localization effects, respectively.
This is attributed to the Berry phase difference between time-reversed trajectories circulating 
the Fermi sphere that encloses the monopole charges. From standard Feynman diagram calculations, we further show that the weak-field magnetoconductivity at low temperatures is proportional to $+\sqrt B$ in double-Weyl semimetals and $-\sqrt{B}$ in single-Weyl semimetals, respectively, which could be verified experimentally.
\end{abstract}

\maketitle

\emph{Introduction} - Topological Weyl semimetal is a new topological state of matter \cite{Weyl29,Nielsen83plb,Volovik,Wan11prb,Xu11prl,Burkov11prl,Yang11prb,halasz2012,zhang2014a, liu2014, weng2015,huang2015,
hirayama2015,soluyanov2015,ruan2015}, in which the conduction and valence energy bands touch nontrivially at discrete momentum points, dubbed Weyl nodes. Remarkably, each Weyl node acts as a magnetic monopole in momentum space. Because the sum of monopole charges of all Weyl nodes in the Brillouin zone is zero, Weyl nodes must appear in pairs, namely the fermion-doubling theorem \cite{Nielsen83plb}. Within each pair, two Weyl nodes carry opposite monopole charges of $\mathcal{N}$ and $-\mathcal{N}$, respectively. Depending on $\mathcal{N}=1$ or $2$, the topological semimetals are referred to as single-Weyl semimetal and double-Weyl semimetal \cite{Xu11prl,Fang12prl,Guan2015prl,Huang15SrSi2}, respectively. The dispersion of the single-Weyl semimetal is linear in three dimensions while the double-Weyl semimetal usually is linear in one and quadratic in the other two dimensions.
Identifying the positions and monopole charges of Weyl nodes is crucial to qualitatively characterize a topological Weyl semimetal and understand the associated novel physical properties.

The monopole charge of a Weyl node also manifests as the number of stable surface Fermi arcs connecting paired Weyl nodes \cite{Wan11prb,Fang12prl}, which can be directly measured in the angle-resolved photoemission spectroscopy (ARPES) experiments \cite{lv2015,xu2015,yang2015,xu2015a,lv2015a}.
As an unambiguous measurement of the surface Fermi arcs in ARPES remains highly challenging, the quantum transport may provide an alternative approach to detect the signature of the monopole charges. For single-Weyl semimetals with $\mathcal{N}=  1$, a $-\sqrt B$ dependence of the magnetoconductivity has been observed near zero magnetic field \cite{Kim13prl,Li14arXiv,HuangXC15prx,ZhangCL15arXiv,Binghai15,ong2015,zhouyi15,li2015nc,Li2016}, which is one of the signatures for the weak antilocalization (WAL) effect in 3D systems \cite{Lu15Weyl-Localization}.
The WAL effect arises from the quantum interference correction to the semiclassical conductivity.
There have been increasing experimental efforts to search for topological Weyl semimetals with higher monopole charges such as ${\mathcal N}=  2$ \cite{Xu11prl,Guan2015prl,Huang15SrSi2}.
Nevertheless, the quantum interference effects on the quantum transport phenomena caused by Weyl nodes with higher monopole charges remains unexplored.

In this Rapid Communication, we explicitly demonstrate the general relation between the monopole charge and quantum transport. We will focus on double-Weyl semimetals with $\mathcal{N}=  2$ and single-Weyl semimetals with ${\mathcal N}= 1$. Based on Feynman diagram calculations, we find that Weyl nodes with $\mathcal{N}=  2$ in double-Weyl semimetals lead to a {\it negative} quantum interference correction to the conductivity compared to the {\it positive} correction in single-Weyl semimetals with $\mathcal N= 1$.
In double-Weyl semimetals, the monopole charge of $\mathcal N=  2$ results in a $2\pi$ Berry phase difference between two time reversed scattering loops. Such two loops can interfere constructively to enhance backscattering, leading to a weak localization (WL) correction in the quantum transport. While for single-Weyl semimetals, the phase difference is $\pi$, leading to destructive interference and hence the WAL effect. The qualitative difference in the quantum transport between single- and double-Weyl semimetals not only provides an experimental signature to identify monopole charges in momentum space but also forward our understanding to the quantum transport in 3D topological metals.

\emph{Weyl semimetals and monopole charges} -
The minimal model which can describe both single- and double-Weyl semimetals can be written as
\begin{equation}\label{hamiltonian}
H=
\left[
  \begin{array}{cc}
    \chi v_z \hbar k_z & v_{\Vert}(\hbar k_{+})^\mathcal{N} \\
    v_{\Vert} (\hbar k_{-})^\mathcal{N} & -\chi v_z \hbar k_z \\
  \end{array}
\right],
\end{equation}
where $k_{\pm}=k_x\pm i k_y$, $\chi=\pm 1$ is the valley index, $v_z$ and $v_{\Vert}$ are parameters and assumed to be constants, and momentum $\bk$ is measured from the Weyl nodes.
Here, $\mathcal{N}=1,2$ correspond to single- and double-Weyl semimetal respectively.
The model has a conduction band and a valence band, with the dispersions given by $\pm E_{\bk}$ and $ E_{\bk}= \sqrt{v_z^2 \hbar^2 k_z^2+v_{\Vert}^2 (\hbar^2 k_x^2+\hbar^2 k_y^2)^{\mathcal{N}}}$.
Without loss of generality, we assume that the chemical potential is slightly above the Weyl nodes and the electronic transport is contributed mainly by the
conduction bands throughout the paper. The eigenstate of the conduction band at valley $\chi=+$ is given by
\begin{eqnarray}\label{eigen-states}
 \ket{\bk}=
 \left[\begin{array}{c}
 \cos (\theta/2)\\
 \sin (\theta/2)\exp(-i \mathcal{N} \varphi)\\
 \end{array}\right],
\end{eqnarray}
where 
$\cos \theta\equiv v_z k_z/E_\bk $, and $\tan \varphi\equiv k_y/k_x$.
The eigenstate of the conduction band around valley $\chi=-$ can be found by replacing $\cos(\theta/2)\rightarrow \sin(\theta/2)$ and $\sin(\theta/2)\rightarrow -\cos(\theta/2)$ in Eq. (\ref{eigen-states}).
The monopole charge can be found by integrating the Berry curvature over an arbitrary Fermi sphere $\Sigma$ that encloses the Weyl node,
\begin{equation}\label{charge}
\frac{1}{2\pi} \int_{\Sigma} d \bm{S}\cdot \mathbf{\Omega}=\pm \mathcal{N},
\end{equation}
with $\pm$ for the $\pm$ valleys, the Berry curvature \cite{Xiao10rmp} $\mathbf{\Omega}=\nabla \times \mathbf{A}$, and $\mathbf{A}=(A_{\theta},A_{\varphi})$ is the Berry connection given by
$A_{\theta}=\bra{\bk}i \partial_{\theta}|\bk \rangle=0$ and $ A_{\varphi}=\bra{\bk}i \partial_{\varphi}|\bk \rangle =\mathcal{N} \sin^2 (\theta/2)$.

\begin{figure}[tb]
\centering
\includegraphics[width=0.5\textwidth]{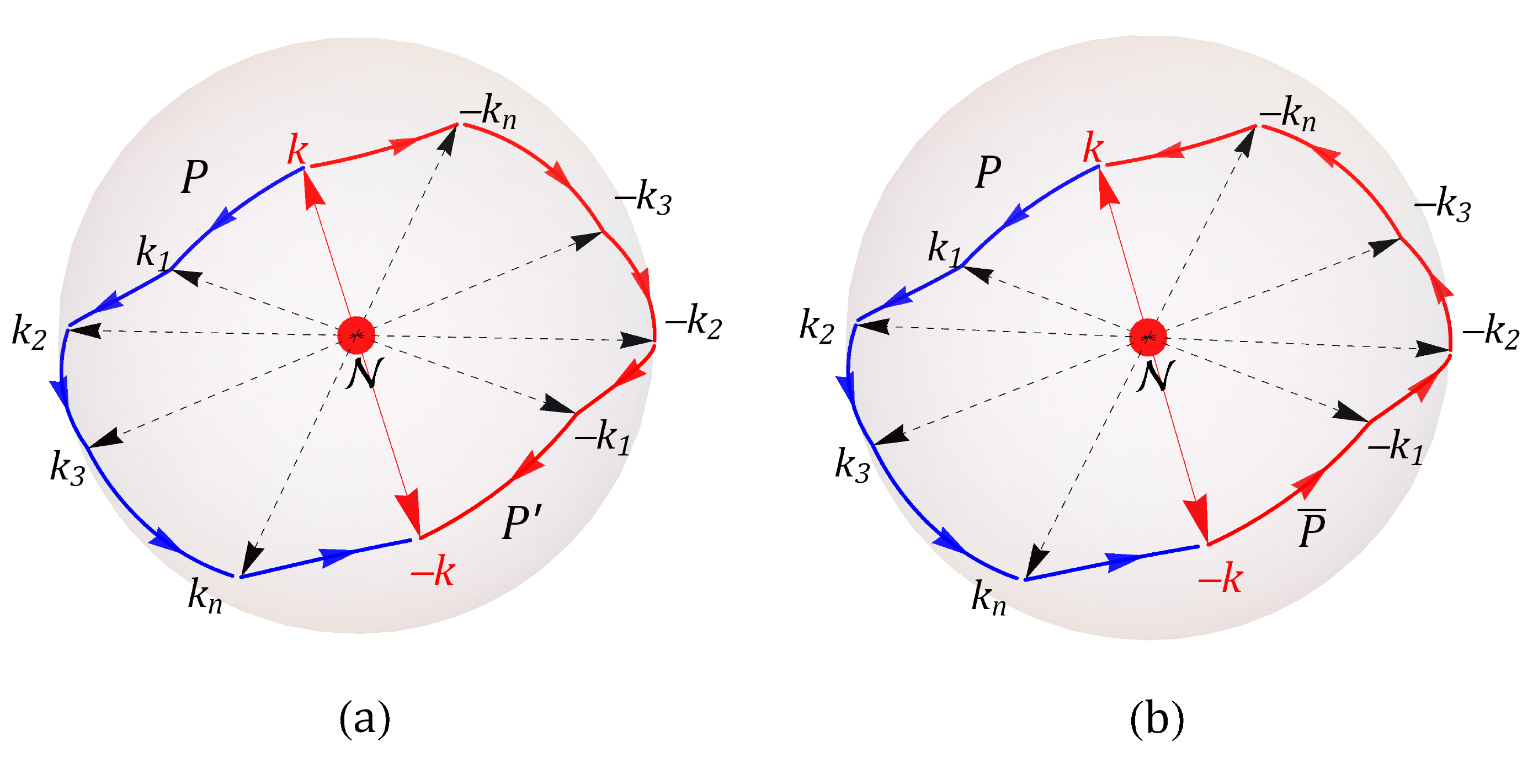}
\protect\caption{The Fermi sphere in momentum space for a three-dimensional topological semimetal, where the dot located at the origin represents a monopole charge of $\mathcal{N}$.
(a) $P$ denotes a generic backscattering from the wave vector $\bk$ to $-\bk$ via intermediate states labeled as ($\bk_1,\bk_2,...,\bk_n$). $P'$ stands for the time-reversal counterpart of $P$.
(b) The phase difference between $P$ and $P'$ is equivalent to the Berry phase circulating around the loop $\mathcal{C}=P+\bar{P}$.}
\label{Fig:loop}
\end{figure}

\emph{Monopole charge and quantum interference} -
In a disordered metal, the backscattering from a state with a wave vector of $\bk$ to $-\bk$ can be achieved by successive scattering via intermediate states. At sufficiently low temperatures, electrons can be scattered by many times but still maintain their phase coherence. In this quantum diffusive regime, the electric conductivity may acquire an extra correction from the quantum interference between the time-reversed scattering paths, leading to the WL or WAL effect \cite{Lee85rmp}. We first focus on the Fermi sphere in one valley and assume no intervalley scattering. For each path [labelled as $P$ in Fig.~\ref{Fig:loop}(a)] connecting successive intermediate states of the backscattering from $\bk$ to $-\bk$ on the Fermi sphere, which encompasses the monopole charge at the origin, there exists a corresponding time-reversal counterpart $P'$. The quantum interference is determined by the phase difference between the two time-reversed paths $P$ and $P'$, which is equivalent to the Berry phase accumulated along the loop formed by $P$ together with $\bar P\equiv -P'$, namely the corresponding path from $-\bk$ to $\bk$, as shown in Fig. \ref{Fig:loop}(b).

The quantum interference correction then depends on the geometric phase, i.e., the Berry phase \cite{Xiao10rmp}, collected by electrons after circulating the loop $\mathcal{C}\equiv P+\bar P$.
The Berry phase can be found by a loop integral of the Berry connection around $\mathcal{C}$.
Remarkably, this Berry phase depends only on the monopole charge, but {\it not} on the specific shape of the loop (see the rigorous proof in \cite{Supp-Weyl}):
\begin{equation}\label{charge_berry}
\gamma =\oint_{\mathcal{C}} d \bm{\ell} \cdot \bA= \pi \mathcal{N}.
\end{equation}
For double-Weyl semimetals, the monopole charges $\mathcal{N}=2$ and the Berry phase is then $2\pi$. With the $ 2\pi$ Berry phase, the time-reversed scattering loops interfere constructively, leading to the weak localization effect.
However, for single-Weyl semimetals, the monopole charge is $\mathcal{N}= 1$ and the Berry phase is $\pi$, which gives rise to the weak antilocalization effect. As the Berry phase is a consequence of the Berry curvature field generated by the monopole charge, we therefore establish a robust connection between the weak (anti)localization effect with the parity of monopole charge $\mathcal{N}$.
The Berry phase argument is consistent with the symmetry classification\cite{Altland97prb},
the single-Weyl	semimetals belong to the symplectic class with a weak antilocalization correction,
while double-Weyl semimetals correspond to the orthogonal class with a weak localization correction.

\begin{figure}[tb]
\centering
\includegraphics[width=0.25\textwidth]{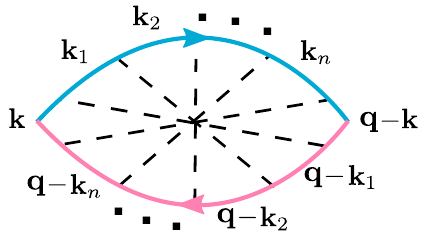}
\protect\caption{The maximally crossed Feynman diagram that describes the quantum interference between the time-reversed scattering trajectories in Fig. \ref{Fig:loop} as $\mathbf{q}\rightarrow 0$. The arrowed solid and dashed lines denote the Green functions and impurity scattering, respectively. This kind of diagrams can give the quantum interference correction to the conductivity \cite{Hikami80ptp,Shon98jpsj,McCann06prl}.
A negative (positive) correction corresponds to the weak (anti)localization effect, with the sign sensitive to the parity of the monopole charge.}
\label{Fig:diagram}
\end{figure}

\emph{Feynman Diagram calculations} -
We now verify the above argument of quantum interference correction to conductivity in Weyl semimetals by the standard Feynman diagram calculations. 
The correction can be evaluated by calculating the maximally crossed diagrams, one of which is shown in Fig.~\ref{Fig:diagram}. In this diagram, the segments of the arrow lines represent the intermediate states in the backscattering, and the dashed lines represent the correlation between the time-reversed scattering processes. The core calculation of the maximally crossed diagrams can be formulated into the particle-particle correlation, known as the cooperon.
The cooperon of the double-Weyl semimetal is found to be (see \cite{Supp-Weyl} for details)
\begin{equation}\label{eq:cooperon2}
\Gamma _{\bk_1,\bk_2}\approx \frac{\hbar }{2\pi N_F \tau ^2}\frac{e^{i2(\varphi_2-\varphi_1)}}{ D_1\left(q_x^2+q_y^2\right)+D_2 q_z^2},
\end{equation}
where $\bq=\bk_1+\bk_2$ is the cooperon wave vector, $\mathbf{k}_1$ and $\mathbf{k}_2$ are the wave vectors of incoming and outgoing states, respectively,
$\varphi_1$ and $\varphi_2$ are the azimuth angles of corresponding wave vectors,
$D_1= 8 \tau E_F v_{\Vert} /3 \pi $ and $D_2= \tau v_z^2$ are the diffusion coefficients,
$N_F$ is the density of states, and $\tau$ is the momentum relaxation time.
In contrast, the cooperon of the single-Weyl semimetal is known to take the form \cite{Lu15Weyl-Localization}
\begin{equation}\label{eq:cooperon1}
\Gamma_{\mathbf{k}_1,\mathbf{k}_2}
\approx \frac{\hbar}{2\pi N_F \tau^2} \frac{1}{ D q^2}e^{i(\varphi_2-\varphi_1)},
\end{equation}
where the diffusion coefficient $D=v^2_F \tau/2$
\footnote{We only give the result for isotropic single-Weyl semimetals with $v_F=v_z=v_{\Vert}$; this simplification does not change any qualitative results with respect to quantum interference correction.}. Note the main difference between Eqs. \eqref{eq:cooperon2} and \eqref{eq:cooperon1} lies in the phase factor involving $\varphi_2-\varphi_1$, which originates from different eigenstates of Weyl semimetals with different monopole charges.

As $\bq\to 0$, i.e., $\bk_1=-\bk_2$, the cooperon becomes divergent and becomes the most dominant contribution to the backscattering. In this limit, $\varphi_2= \varphi_1+ \pi$
\footnote{We have carried out a coordinate transformation in deriving these results, where $\hbar k_x=\sqrt{k \sin \theta }\cos \varphi, \hbar k_y=\sqrt{k \sin \theta }\sin \varphi, 2\hbar m v k_z =k \cos \theta $, $-\bk$ is obtained by setting $\varphi \to \varphi+ \pi$ and $\theta\to \pi-\theta$.}.
Then, for the double-Weyl semimetal,
\begin{eqnarray}\label{eq:coop2}
\Gamma_{\bk,\bq-\bk}&\approx &+\frac{\hbar }{2\pi N_F \tau ^2}\frac{1}{ D_1\left(q_x^2+q_y^2\right)+D_2 q_z^2},
\end{eqnarray}
and for the single-Weyl semimetal,
\begin{eqnarray}\label{eq:coop1}
\Gamma_{\bk,\bq-\bk}
&\approx &-\frac{\hbar}{2\pi N_F \tau^2} \frac{1}{ D q^2}.	
\end{eqnarray}
Note the different signs in Eqs. \eqref{eq:coop2} and \eqref{eq:coop1}, which correspond to the WL and WAL effects, respectively.
This is a direct consequence of different phase factors in the wavefunctions, generated by different monopole charges in double- and single-Weyl semimetals.
In other words, a connection is therefore firmly established between the \emph{parity} of monopole charge $\mathcal{N}$ and the sign of the quantum interference correction, with odd and even parity giving rise to WAL and WL, respectively. This is the main result of the paper.

\emph{Weak-localization conductivity} - 
From the obtained cooperon above, it is straightforward to compute the quantum interference corrections to the Drude conductivity (see \cite{Supp-Weyl} for details). For double-Weyl semimetals, the quantum interference corrections to the Drude conductivity are intrinsically anisotropic because of the anisotropic dispersions in double-Weyl semimetals
\begin{eqnarray}\label{sigma-qi}
\sigma_{zz}^\textrm{qi}&=&-\frac{9e^2}{4 \pi^2 \hbar }\frac{v_z}{2}\sqrt{\frac{1}{2 v_{\Vert}}
\frac{1}{\pi E_F}}\left(\frac{1}{\ell_z}-\frac{1}{\ell_{\phi}}\right),\nonumber\\
\sigma_{xx}^\textrm{qi}&=&-\frac{e^2}{4 \pi^2 \hbar}\frac{2}{v_z}\sqrt{2 v_{\Vert}\pi E_F}
\left(\frac{1}{\ell_x}-\frac{1}{\ell_{\phi}}\right),
\end{eqnarray}
where $\ell_i$ is the mean free path and $\ell_{\phi}$ is the phase coherence length.
The mean free path $\ell_i$ is defined as $\ell_i=\sqrt{D_i \tau}$.
In the quantum diffusive regime, $\ell_i$ is much shorter than $\ell_\phi$, so the correction is negative in all directions.
This negative correction is one of the signatures of the WL effect, consistent with the above Berry-phase argument of monopole charge.
Note that this negative correction due to the WL effect is not divergent, as expected in 3D.
Moreover, the total conductivity $\sigma=\sigma^\textrm{sc}+\sigma^\textrm{qi}$, where $\sigma^\textrm{sc}$ is the semiclassical Drude conductivity, is still finite.
From $\ell_i=v_i \tau$ and $\tau\propto 1/E_F$, we obtain that $\sigma_{zz}^\textrm{qi}\propto -E_F^{1/2}$ and $\sigma^\textrm{qi}_{xx}\propto -E_F^{3/2}$ at low temperatures. The semiclassical conductivity is found to be $\sigma_{zz}^\textrm{sc}=e^2N_F v_z^2 \tau \propto E_F^0$
and $\sigma_{xx}^\textrm{sc}=(8/3 \pi ) e^2 \tau N_F E_F v_{\Vert}\propto E_F$, where the density of states at the Fermi level per valley is $N_F= E_F/8\pi \hbar^3 v_z v_{\Vert}$.

\emph{Weak localization induced magnetoconductivity} 
- The negative quantum interference correction in Eq. (\ref{sigma-qi}) can be destroyed by a small magnetic field, giving rise to a positive magnetoconductivity as another signature of the weak localization in double-Weyl semimetals.
The magnetoconductivity is anisotropic, depending on the field direction. When the field is applied along the $z$ direction, the quantum interference correction of the conductivity is found as
\begin{eqnarray}\label{digamma}
&&\sigma^\textrm{qi}_{zz}(B)=-\frac{3e^2 \tau v_z^2}{8 \hbar \pi^2}\int_{0}^{1/\ell_z}d q_z \nonumber\\
&&\times \left[\psi\left(\frac{\ell_{B}^2}{ \ell^2_z}+\ell_{B }^2 q_z^2+\frac{1}{2}\right)-\psi\left(\frac{\ell_{B}^2}{\ell^2_{\phi}}+\ell_{B }^2 q_z^2+\frac{1}{2}\right)\right],
\end{eqnarray}
where $\psi$ is the digamma function and $\ell_{B}\equiv \sqrt{ \hbar/4eD_1 B}$ is the effective magnetic length in the $z$ direction. When the field is applied in the $x$-$y$ plane, the conductivity is given by the same formula as Eq. \eqref{digamma} but with $\ell_{B}^2=  \hbar/4eB\sqrt{D_1 D_2}$. The magnetoconductivity is defined as $\delta\sigma^\textrm{qi}_{zz}(B)=\sigma^\textrm{qi}_{zz}(B)-\sigma^\textrm{qi}_{zz}(0)$. 
In the limit of $\ell_{\phi}\gg \ell_B \gg \ell_z$, which can be approached at low temperatures, the magnetoconductivity $\delta \sigma^\textrm{qi}_{zz}(B) \propto \sqrt{B}$. In the limit of $\ell_B\gg \ell_\phi$ and $\ell_B\gg \ell_z$, $\delta \sigma^\textrm{qi}_{zz}(B) \propto B^2$. As a good approximation, the weak localization induced magnetoconductivity in double-Weyl semimetals can be fitted as
\begin{eqnarray}\label{fitting-formula}
\delta\sigma^\textrm{qi}_{zz} &=& C^\textrm{qi}_1\frac{B^2\sqrt{B}}{B_c^2+B^2}+C^\textrm{qi}_2\frac{B_c^2 B^2}{B_c^2+B^2},
\end{eqnarray}
where the fitting parameters $C^\textrm{qi}_{1}$ and $C^\textrm{qi}_{2}$ are positive and the critical field $B_c$ is related to the phase coherence length $\ell_\phi$ according to $B_c\sim \hbar/e\ell_\phi^2$. Empirically, the phase coherence length becomes longer with decreasing temperature and can be written as $\ell_\phi \sim T^{-p/2}$; then $B_c \sim   T^p$, where $p$ is positive and determined by decoherence mechanisms such as electron-electron interaction ($p=3/2$) or electron-phonon interaction ($p=3$). At high temperatures, $\ell_\phi\rightarrow 0$; thus, $B_c \rightarrow \infty$ and we have $\delta\sigma^\textrm{qi}_{zz}\propto B^2$. At low temperatures, $\ell_\phi\rightarrow \infty$; then $B_c=0$ and we have $\delta\sigma^\textrm{qi}_{zz}\propto\sqrt{B}$.

\emph{Chiral-anomaly induced magnetoconductivity} -
Another mechanism that can also contribute a positive magnetoconductivity is the chiral anomaly.
The nontrivial momentum-space Berry curvature in Weyl semimetals can induce an anomalous velocity which can couple with external magnetic fields \cite{Xiao10rmp}, leading to a positive magnetoconductivity when the electric current is in parallel with magnetic fields \cite{Son13prb,Burkov14prl-chiral}.
In single-Weyl semimetals at weak fields, this ``chiral anomaly'' induced magnetoconductivity is found to be proportional to $ B^2 $ \cite{Son13prb,Burkov14prl-chiral} or $B$ \cite{ZhangSB15arXiv}, depending on the Fermi energy.
As a semiclassical effect, the chiral anomaly is not as sensitive to phase coherence and temperature
as the weak localization effect is,
so a similar positive magnetoconductivity is also expected in double-Weyl semimetals, without sign reversing, although the monopole charge is doubled.
In addition, the chiral anomaly has a strong angle dependence, becomes prominent only in parallel magnetic fields. In contrast, the weak localization happens in all directions. These properties distinguish the positive magnetoconductivity of the chiral anomaly from that of the weak localization effect. In perpendicular magnetic fields, the Lorentz force gives rise to the classical negative magnetoconductivity that is proportional to $-\mu^2 B^2$,
where the mobility $\mu=2e D/E_F$ in double-Weyl semimetals.
Because of the functional relations, at sufficiently low temperatures the WL effect induced $\sqrt B$ magnetoconductivity always overwhelms the magnetoconductivity from the chiral anomaly and Lorentz force. For a comparison between single- and double-Weyl semimetals, we summarize the key results in Table \ref{tab:2D-3D}.

\begin{table}[tb]
\caption{Comparison between 3D single- and double-Weyl semimetals.
$\pm E_{\bk}$ is the dispersion relation of the conduction and valence bands,
$N(E)$ is the density of states;
$\delta\sigma^{qi}(B)$ is the weak (anti)localization magnetoconductivity when $\ell_{\phi}\gg\ell_{B}$. $\pm\mathcal{N}$ is the monopole charge.}
\label{tab:2D-3D}%
\begin{ruledtabular}
\begin{tabular}{ccc}
 & 3D single-Weyl & 3D double-Weyl \\
\hline
 $E_{\bk}$ &  $  \hbar \sqrt{v_{\Vert}^2  (k_x^2+k_y^2)+ v_z^2 k_z^2}$  & $ \hbar \sqrt{v_{\Vert}^2 \hbar^2 (k_x^2+k_y^2)^2+ v_z^2 k_z^2}$ \\
$N(E)$ &  $E^2/ 2\pi^2 v^2_{\Vert}v_z \hbar^3 $  & $ E/ 8\pi v_z v_{\Vert}\hbar^3 $  \\
$\delta\sigma^{qi}(B)$  & $-\sqrt{B}$  & $\sqrt{B}$  \tabularnewline
$\mathcal{N}$  & $ 1$  & $ 2$  \tabularnewline
\end{tabular}
\end{ruledtabular}
\end{table}

\emph{Effects of intervalley scattering} -
For single-Weyl semimetals, in the limit where intervalley scattering dominates, the quantum interference correction can be negative, i.e., a crossover to the weak localization \cite{Lu15Weyl-Localization}, which is a result of the cancellation of $\pm \pi$ Berry phase when circulating monopoles with opposite charges.
Similar phenomena were also studied in 2D graphene \cite{Suzuura02prl,McCann06prl,Tikhonenko09prl,Ando98jpsj}.
However, in double-Weyl semimetals, even if intervalley scattering is considered where the electron scattering trajectory encloses the nodes of double-Weyl semimetals with opposite monopole charges, the weak localization is still robust since the total phase acquired by circulating the two monopoles is still an integer multiple of $2\pi$. 
The crossover can be also understood from the symmetry classification\cite{Altland97prb},
with $\pm \pi$ ($\pm 2 \pi$) Berry phase corresponding to the symplectic (orthogonal) class.
In the limit of weak intervalley scattering, the qualitatively different quantum interference corrections can distinguish the Weyl nodes with different monopole charges.

\emph{Conclusions and discussions} - To summarize, we have presented an intuitive picture revealing the connection between quantum interference correction to the conductivity and monopole charge in three-dimensional Weyl semimetals. The picture is verified by standard Feynman diagram calculations. It explains the weak antilocalization observed in single-Weyl semimetals \cite{Kim13prl,Li14arXiv,HuangXC15prx,ZhangCL15arXiv} and predicts the weak localization in double-Weyl semimetals.
Experimentally, the quantum interference correction can manifest through magnetoconductivity. In the zero temperature limit, the magnetoconductivity $\propto \mp \sqrt{B}$ for single- and double-Weyl semimetals, respectively. So far, there have been several materials, e.g., $\rm{HgCr_2Se_4}$ \cite{Xu11prl, Fang12prl} and $\rm{SrSi_2}$ \cite{Huang15SrSi2} proposed as candidates for double-Weyl semimetals.

It is worthwhile to mention that, if weak higher order terms which may break the time-reversal symmetry are included in the $\bk\cdot \bp $ Hamiltonian, the quantum interference correction will be slightly smaller as pointed out in Ref.~\cite{Ando98jpsj}. Even though interactions can also contribute to conductivity,
interaction-induced magnetoconductivity usually has much weaker magnetic field dependence compared to the one induced by the quantum interference \cite{Lee85rmp,Lu15Weyl-Localization,Lu14prl}; it is expected that our results above are robust if weak interactions are included.

{\it Acknowledgment.} We would like to thank Zi-Xiang Li and Shao-Kai Jian for stimulating discussions. This work was supported by the NSFC under Grant No. 11474175 (XD and HY) and 11574127 (HZL), the National Thousand-Young-Talents Program (HZL and HY), and by the Research
Grant Council, University Grants Committee, Hong Kong under Grant
No. 17303714.


\begin{thebibliography}{49}%
\makeatletter
\providecommand \@ifxundefined [1]{%
 \@ifx{#1\undefined}
}%
\providecommand \@ifnum [1]{%
 \ifnum #1\expandafter \@firstoftwo
 \else \expandafter \@secondoftwo
 \fi
}%
\providecommand \@ifx [1]{%
 \ifx #1\expandafter \@firstoftwo
 \else \expandafter \@secondoftwo
 \fi
}%
\providecommand \natexlab [1]{#1}%
\providecommand \enquote  [1]{``#1''}%
\providecommand \bibnamefont  [1]{#1}%
\providecommand \bibfnamefont [1]{#1}%
\providecommand \citenamefont [1]{#1}%
\providecommand \href@noop [0]{\@secondoftwo}%
\providecommand \href [0]{\begingroup \@sanitize@url \@href}%
\providecommand \@href[1]{\@@startlink{#1}\@@href}%
\providecommand \@@href[1]{\endgroup#1\@@endlink}%
\providecommand \@sanitize@url [0]{\catcode `\\12\catcode `\$12\catcode
  `\&12\catcode `\#12\catcode `\^12\catcode `\_12\catcode `\%12\relax}%
\providecommand \@@startlink[1]{}%
\providecommand \@@endlink[0]{}%
\providecommand \url  [0]{\begingroup\@sanitize@url \@url }%
\providecommand \@url [1]{\endgroup\@href {#1}{\urlprefix }}%
\providecommand \urlprefix  [0]{URL }%
\providecommand \Eprint [0]{\href }%
\providecommand \doibase [0]{http://dx.doi.org/}%
\providecommand \selectlanguage [0]{\@gobble}%
\providecommand \bibinfo  [0]{\@secondoftwo}%
\providecommand \bibfield  [0]{\@secondoftwo}%
\providecommand \translation [1]{[#1]}%
\providecommand \BibitemOpen [0]{}%
\providecommand \bibitemStop [0]{}%
\providecommand \bibitemNoStop [0]{.\EOS\space}%
\providecommand \EOS [0]{\spacefactor3000\relax}%
\providecommand \BibitemShut  [1]{\csname bibitem#1\endcsname}%
\let\auto@bib@innerbib\@empty
\bibitem [{\citenamefont {Weyl}(1929)}]{Weyl29}%
  \BibitemOpen
  \bibfield  {author} {\bibinfo {author} {\bibfnamefont {H.}~\bibnamefont
  {Weyl}},\ }\href@noop {} {\bibfield  {journal} {\bibinfo  {journal} {Z.
  Phys.}\ }\textbf {\bibinfo {volume} {56}},\ \bibinfo {pages} {330} (\bibinfo
  {year} {1929})}\BibitemShut {NoStop}%
\bibitem [{\citenamefont {Nielsen}\ and\ \citenamefont
  {Ninomiya}(1983)}]{Nielsen83plb}%
  \BibitemOpen
  \bibfield  {author} {\bibinfo {author} {\bibfnamefont {H.~B.}\ \bibnamefont
  {Nielsen}}\ and\ \bibinfo {author} {\bibfnamefont {M.}~\bibnamefont
  {Ninomiya}},\ }\href {\doibase
  http://dx.doi.org/10.1016/0370-2693(83)91529-0} {\bibfield  {journal}
  {\bibinfo  {journal} {Physics Letters B}\ }\textbf {\bibinfo {volume}
  {130}},\ \bibinfo {pages} {389 } (\bibinfo {year} {1983})}\BibitemShut
  {NoStop}%
\bibitem [{\citenamefont {Volovik}(2009)}]{Volovik}%
  \BibitemOpen
  \bibfield  {author} {\bibinfo {author} {\bibfnamefont {G.~E.}\ \bibnamefont
  {Volovik}},\ }\href@noop {} {\emph {\bibinfo {title} {The universe in a
  helium droplet}}}\ (\bibinfo  {publisher} {Oxford University Press},\
  \bibinfo {year} {2009})\BibitemShut {NoStop}%
\bibitem [{\citenamefont {Wan}\ \emph {et~al.}(2011)\citenamefont {Wan},
  \citenamefont {Turner}, \citenamefont {Vishwanath},\ and\ \citenamefont
  {Savrasov}}]{Wan11prb}%
  \BibitemOpen
  \bibfield  {author} {\bibinfo {author} {\bibfnamefont {X.}~\bibnamefont
  {Wan}}, \bibinfo {author} {\bibfnamefont {A.~M.}\ \bibnamefont {Turner}},
  \bibinfo {author} {\bibfnamefont {A.}~\bibnamefont {Vishwanath}}, \ and\
  \bibinfo {author} {\bibfnamefont {S.~Y.}\ \bibnamefont {Savrasov}},\ }\href
  {\doibase 10.1103/PhysRevB.83.205101} {\bibfield  {journal} {\bibinfo
  {journal} {Phys. Rev. B}\ }\textbf {\bibinfo {volume} {83}},\ \bibinfo
  {pages} {205101} (\bibinfo {year} {2011})}\BibitemShut {NoStop}%
\bibitem [{\citenamefont {Xu}\ \emph {et~al.}(2011)\citenamefont {Xu},
  \citenamefont {Weng}, \citenamefont {Wang}, \citenamefont {Dai},\ and\
  \citenamefont {Fang}}]{Xu11prl}%
  \BibitemOpen
  \bibfield  {author} {\bibinfo {author} {\bibfnamefont {G.}~\bibnamefont
  {Xu}}, \bibinfo {author} {\bibfnamefont {H.~M.}\ \bibnamefont {Weng}},
  \bibinfo {author} {\bibfnamefont {Z.~J.}\ \bibnamefont {Wang}}, \bibinfo
  {author} {\bibfnamefont {X.}~\bibnamefont {Dai}}, \ and\ \bibinfo {author}
  {\bibfnamefont {Z.}~\bibnamefont {Fang}},\ }\href {\doibase
  10.1103/PhysRevLett.107.186806} {\bibfield  {journal} {\bibinfo  {journal}
  {Phys. Rev. Lett.}\ }\textbf {\bibinfo {volume} {107}},\ \bibinfo {pages}
  {186806} (\bibinfo {year} {2011})}\BibitemShut {NoStop}%
\bibitem [{\citenamefont {Burkov}\ and\ \citenamefont
  {Balents}(2011)}]{Burkov11prl}%
  \BibitemOpen
  \bibfield  {author} {\bibinfo {author} {\bibfnamefont {A.~A.}\ \bibnamefont
  {Burkov}}\ and\ \bibinfo {author} {\bibfnamefont {L.}~\bibnamefont
  {Balents}},\ }\href {\doibase 10.1103/PhysRevLett.107.127205} {\bibfield
  {journal} {\bibinfo  {journal} {Phys. Rev. Lett.}\ }\textbf {\bibinfo
  {volume} {107}},\ \bibinfo {pages} {127205} (\bibinfo {year}
  {2011})}\BibitemShut {NoStop}%
\bibitem [{\citenamefont {Yang}\ \emph {et~al.}(2011)\citenamefont {Yang},
  \citenamefont {Lu},\ and\ \citenamefont {Ran}}]{Yang11prb}%
  \BibitemOpen
  \bibfield  {author} {\bibinfo {author} {\bibfnamefont {K.~Y.}\ \bibnamefont
  {Yang}}, \bibinfo {author} {\bibfnamefont {Y.~M.}\ \bibnamefont {Lu}}, \ and\
  \bibinfo {author} {\bibfnamefont {Y.}~\bibnamefont {Ran}},\ }\href {\doibase
  10.1103/PhysRevB.84.075129} {\bibfield  {journal} {\bibinfo  {journal} {Phys.
  Rev. B}\ }\textbf {\bibinfo {volume} {84}},\ \bibinfo {pages} {075129}
  (\bibinfo {year} {2011})}\BibitemShut {NoStop}%
\bibitem [{\citenamefont {Hal\'asz}\ and\ \citenamefont
  {Balents}(2012)}]{halasz2012}%
  \BibitemOpen
  \bibfield  {author} {\bibinfo {author} {\bibfnamefont {G.~B.}\ \bibnamefont
  {Hal\'asz}}\ and\ \bibinfo {author} {\bibfnamefont {L.}~\bibnamefont
  {Balents}},\ }\href {\doibase 10.1103/PhysRevB.85.035103} {\bibfield
  {journal} {\bibinfo  {journal} {Phys. Rev. B}\ }\textbf {\bibinfo {volume}
  {85}},\ \bibinfo {pages} {035103} (\bibinfo {year} {2012})}\BibitemShut
  {NoStop}%
\bibitem [{\citenamefont {Zhang}\ \emph {et~al.}(2014)\citenamefont {Zhang},
  \citenamefont {Wang}, \citenamefont {Xu}, \citenamefont {Xu},\ and\
  \citenamefont {Zhang}}]{zhang2014a}%
  \BibitemOpen
  \bibfield  {author} {\bibinfo {author} {\bibfnamefont {H.}~\bibnamefont
  {Zhang}}, \bibinfo {author} {\bibfnamefont {J.}~\bibnamefont {Wang}},
  \bibinfo {author} {\bibfnamefont {G.}~\bibnamefont {Xu}}, \bibinfo {author}
  {\bibfnamefont {Y.}~\bibnamefont {Xu}}, \ and\ \bibinfo {author}
  {\bibfnamefont {S.-C.}\ \bibnamefont {Zhang}},\ }\href
  {http://journals.aps.org/prl/abstract/10.1103/PhysRevLett.112.096804}
  {\bibfield  {journal} {\bibinfo  {journal} {Phys. Rev. Lett.}\ }\textbf
  {\bibinfo {volume} {112}},\ \bibinfo {pages} {096804} (\bibinfo {year}
  {2014})}\BibitemShut {NoStop}%
\bibitem [{\citenamefont {Liu}\ and\ \citenamefont
  {Vanderbilt}(2014)}]{liu2014}%
  \BibitemOpen
  \bibfield  {author} {\bibinfo {author} {\bibfnamefont {J.}~\bibnamefont
  {Liu}}\ and\ \bibinfo {author} {\bibfnamefont {D.}~\bibnamefont
  {Vanderbilt}},\ }\href
  {http://journals.aps.org/prb/abstract/10.1103/PhysRevB.90.155316} {\bibfield
  {journal} {\bibinfo  {journal} {Phys. Rev. B}\ }\textbf {\bibinfo {volume}
  {90}},\ \bibinfo {pages} {155316} (\bibinfo {year} {2014})}\BibitemShut
  {NoStop}%
\bibitem [{\citenamefont {Weng}\ \emph {et~al.}(2015)\citenamefont {Weng},
  \citenamefont {Fang}, \citenamefont {Fang}, \citenamefont {Bernevig},\ and\
  \citenamefont {Dai}}]{weng2015}%
  \BibitemOpen
  \bibfield  {author} {\bibinfo {author} {\bibfnamefont {H.}~\bibnamefont
  {Weng}}, \bibinfo {author} {\bibfnamefont {C.}~\bibnamefont {Fang}}, \bibinfo
  {author} {\bibfnamefont {Z.}~\bibnamefont {Fang}}, \bibinfo {author}
  {\bibfnamefont {B.~A.}\ \bibnamefont {Bernevig}}, \ and\ \bibinfo {author}
  {\bibfnamefont {X.}~\bibnamefont {Dai}},\ }\href
  {http://journals.aps.org/prx/abstract/10.1103/PhysRevX.5.011029} {\bibfield
  {journal} {\bibinfo  {journal} {Phys. Rev. X}\ }\textbf {\bibinfo {volume}
  {5}},\ \bibinfo {pages} {011029} (\bibinfo {year} {2015})}\BibitemShut
  {NoStop}%
\bibitem [{\citenamefont {Huang}\ \emph
  {et~al.}(2015{\natexlab{a}})\citenamefont {Huang}, \citenamefont {Xu},
  \citenamefont {Belopolski}, \citenamefont {Lee}, \citenamefont {Chang},
  \citenamefont {Wang}, \citenamefont {Alidoust}, \citenamefont {Bian},
  \citenamefont {Neupane}, \citenamefont {Zhang}, \citenamefont {Jia},
  \citenamefont {Bansil}, \citenamefont {Lin},\ and\ \citenamefont
  {Hasan}}]{huang2015}%
  \BibitemOpen
  \bibfield  {author} {\bibinfo {author} {\bibfnamefont {S.-M.}\ \bibnamefont
  {Huang}}, \bibinfo {author} {\bibfnamefont {S.-Y.}\ \bibnamefont {Xu}},
  \bibinfo {author} {\bibfnamefont {I.}~\bibnamefont {Belopolski}}, \bibinfo
  {author} {\bibfnamefont {C.-C.}\ \bibnamefont {Lee}}, \bibinfo {author}
  {\bibfnamefont {G.}~\bibnamefont {Chang}}, \bibinfo {author} {\bibfnamefont
  {B.}~\bibnamefont {Wang}}, \bibinfo {author} {\bibfnamefont {N.}~\bibnamefont
  {Alidoust}}, \bibinfo {author} {\bibfnamefont {G.}~\bibnamefont {Bian}},
  \bibinfo {author} {\bibfnamefont {M.}~\bibnamefont {Neupane}}, \bibinfo
  {author} {\bibfnamefont {C.}~\bibnamefont {Zhang}}, \bibinfo {author}
  {\bibfnamefont {S.}~\bibnamefont {Jia}}, \bibinfo {author} {\bibfnamefont
  {A.}~\bibnamefont {Bansil}}, \bibinfo {author} {\bibfnamefont
  {H.}~\bibnamefont {Lin}}, \ and\ \bibinfo {author} {\bibfnamefont {M.~Z.}\
  \bibnamefont {Hasan}},\ }\href
  {http://www.nature.com/ncomms/2015/150612/ncomms8373/full/ncomms8373.html}
  {\bibfield  {journal} {\bibinfo  {journal} {Nat. Commun.}\ }\textbf {\bibinfo
  {volume} {6}},\ \bibinfo {pages} {7373} (\bibinfo {year}
  {2015}{\natexlab{a}})}\BibitemShut {NoStop}%
\bibitem [{\citenamefont {Hirayama}\ \emph {et~al.}(2015)\citenamefont
  {Hirayama}, \citenamefont {Okugawa}, \citenamefont {Ishibashi}, \citenamefont
  {Murakami},\ and\ \citenamefont {Miyake}}]{hirayama2015}%
  \BibitemOpen
  \bibfield  {author} {\bibinfo {author} {\bibfnamefont {M.}~\bibnamefont
  {Hirayama}}, \bibinfo {author} {\bibfnamefont {R.}~\bibnamefont {Okugawa}},
  \bibinfo {author} {\bibfnamefont {S.}~\bibnamefont {Ishibashi}}, \bibinfo
  {author} {\bibfnamefont {S.}~\bibnamefont {Murakami}}, \ and\ \bibinfo
  {author} {\bibfnamefont {T.}~\bibnamefont {Miyake}},\ }\href {\doibase
  10.1103/PhysRevLett.114.206401} {\bibfield  {journal} {\bibinfo  {journal}
  {Phys. Rev. Lett.}\ }\textbf {\bibinfo {volume} {114}},\ \bibinfo {pages}
  {206401} (\bibinfo {year} {2015})}\BibitemShut {NoStop}%
\bibitem [{\citenamefont {Soluyanov}\ \emph {et~al.}(2015)\citenamefont
  {Soluyanov}, \citenamefont {Gresch}, \citenamefont {Wang}, \citenamefont
  {Wu}, \citenamefont {Troyer}, \citenamefont {Dai},\ and\ \citenamefont
  {Bernevig}}]{soluyanov2015}%
  \BibitemOpen
  \bibfield  {author} {\bibinfo {author} {\bibfnamefont {A.~A.}\ \bibnamefont
  {Soluyanov}}, \bibinfo {author} {\bibfnamefont {D.}~\bibnamefont {Gresch}},
  \bibinfo {author} {\bibfnamefont {Z.}~\bibnamefont {Wang}}, \bibinfo {author}
  {\bibfnamefont {Q.}~\bibnamefont {Wu}}, \bibinfo {author} {\bibfnamefont
  {M.}~\bibnamefont {Troyer}}, \bibinfo {author} {\bibfnamefont
  {X.}~\bibnamefont {Dai}}, \ and\ \bibinfo {author} {\bibfnamefont {B.~A.}\
  \bibnamefont {Bernevig}},\ }\href
  {http://www.nature.com/nature/journal/v527/n7579/abs/nature15768.html}
  {\bibfield  {journal} {\bibinfo  {journal} {Nature}\ }\textbf {\bibinfo
  {volume} {527}},\ \bibinfo {pages} {495} (\bibinfo {year}
  {2015})}\BibitemShut {NoStop}%
\bibitem [{\citenamefont {Ruan}\ \emph {et~al.}(2016)\citenamefont {Ruan},
  \citenamefont {Jian}, \citenamefont {Yao}, \citenamefont {Zhang},
  \citenamefont {Zhang},\ and\ \citenamefont {Xing}}]{ruan2015}%
  \BibitemOpen
  \bibfield  {author} {\bibinfo {author} {\bibfnamefont {J.}~\bibnamefont
  {Ruan}}, \bibinfo {author} {\bibfnamefont {S.-K.}\ \bibnamefont {Jian}},
  \bibinfo {author} {\bibfnamefont {H.}~\bibnamefont {Yao}}, \bibinfo {author}
  {\bibfnamefont {H.}~\bibnamefont {Zhang}}, \bibinfo {author} {\bibfnamefont
  {S.-C.}\ \bibnamefont {Zhang}}, \ and\ \bibinfo {author} {\bibfnamefont
  {D.}~\bibnamefont {Xing}},\ }\href
  {http://www.nature.com/ncomms/2016/160401/ncomms11136/full/ncomms11136.html}
  {\bibfield  {journal} {\bibinfo  {journal} {Nat. Commun.}\ }\textbf {\bibinfo
  {volume} {7}} (\bibinfo {year} {2016})}\BibitemShut {NoStop}%
\bibitem [{\citenamefont {Fang}\ \emph {et~al.}(2012)\citenamefont {Fang},
  \citenamefont {Gilbert}, \citenamefont {Dai},\ and\ \citenamefont
  {Bernevig}}]{Fang12prl}%
  \BibitemOpen
  \bibfield  {author} {\bibinfo {author} {\bibfnamefont {C.}~\bibnamefont
  {Fang}}, \bibinfo {author} {\bibfnamefont {M.~J.}\ \bibnamefont {Gilbert}},
  \bibinfo {author} {\bibfnamefont {X.}~\bibnamefont {Dai}}, \ and\ \bibinfo
  {author} {\bibfnamefont {B.~A.}\ \bibnamefont {Bernevig}},\ }\href {\doibase
  10.1103/PhysRevLett.108.266802} {\bibfield  {journal} {\bibinfo  {journal}
  {Phys. Rev. Lett.}\ }\textbf {\bibinfo {volume} {108}},\ \bibinfo {pages}
  {266802} (\bibinfo {year} {2012})}\BibitemShut {NoStop}%
\bibitem [{\citenamefont {Guan}\ \emph {et~al.}(2015)\citenamefont {Guan},
  \citenamefont {Lin}, \citenamefont {Yang}, \citenamefont {Shi}, \citenamefont
  {Ren}, \citenamefont {Li}, \citenamefont {Weng}, \citenamefont {Dai},
  \citenamefont {Fang}, \citenamefont {Yan},\ and\ \citenamefont
  {Xiong}}]{Guan2015prl}%
  \BibitemOpen
  \bibfield  {author} {\bibinfo {author} {\bibfnamefont {T.}~\bibnamefont
  {Guan}}, \bibinfo {author} {\bibfnamefont {C.}~\bibnamefont {Lin}}, \bibinfo
  {author} {\bibfnamefont {C.}~\bibnamefont {Yang}}, \bibinfo {author}
  {\bibfnamefont {Y.}~\bibnamefont {Shi}}, \bibinfo {author} {\bibfnamefont
  {C.}~\bibnamefont {Ren}}, \bibinfo {author} {\bibfnamefont {Y.}~\bibnamefont
  {Li}}, \bibinfo {author} {\bibfnamefont {H.}~\bibnamefont {Weng}}, \bibinfo
  {author} {\bibfnamefont {X.}~\bibnamefont {Dai}}, \bibinfo {author}
  {\bibfnamefont {Z.}~\bibnamefont {Fang}}, \bibinfo {author} {\bibfnamefont
  {S.}~\bibnamefont {Yan}}, \ and\ \bibinfo {author} {\bibfnamefont
  {P.}~\bibnamefont {Xiong}},\ }\href {\doibase 10.1103/PhysRevLett.115.087002}
  {\bibfield  {journal} {\bibinfo  {journal} {Phys. Rev. Lett.}\ }\textbf
  {\bibinfo {volume} {115}},\ \bibinfo {pages} {087002} (\bibinfo {year}
  {2015})}\BibitemShut {NoStop}%
\bibitem [{\citenamefont {Huang}\ \emph {et~al.}(2016)\citenamefont {Huang},
  \citenamefont {Xu}, \citenamefont {Belopolski}, \citenamefont {Lee},
  \citenamefont {Chang}, \citenamefont {Chang}, \citenamefont {Wang},
  \citenamefont {Alidoust}, \citenamefont {Bian}, \citenamefont {Neupane} \emph
  {et~al.}}]{Huang15SrSi2}%
  \BibitemOpen
  \bibfield  {author} {\bibinfo {author} {\bibfnamefont {S.-M.}\ \bibnamefont
  {Huang}}, \bibinfo {author} {\bibfnamefont {S.-Y.}\ \bibnamefont {Xu}},
  \bibinfo {author} {\bibfnamefont {I.}~\bibnamefont {Belopolski}}, \bibinfo
  {author} {\bibfnamefont {C.-C.}\ \bibnamefont {Lee}}, \bibinfo {author}
  {\bibfnamefont {G.}~\bibnamefont {Chang}}, \bibinfo {author} {\bibfnamefont
  {T.-R.}\ \bibnamefont {Chang}}, \bibinfo {author} {\bibfnamefont
  {B.}~\bibnamefont {Wang}}, \bibinfo {author} {\bibfnamefont {N.}~\bibnamefont
  {Alidoust}}, \bibinfo {author} {\bibfnamefont {G.}~\bibnamefont {Bian}},
  \bibinfo {author} {\bibfnamefont {M.}~\bibnamefont {Neupane}},  \emph
  {et~al.},\ }\href {http://www.pnas.org/content/113/5/1180} {\bibfield
  {journal} {\bibinfo  {journal} {PNAS}\ }\textbf {\bibinfo {volume} {113}}
  (\bibinfo {year} {2016})}\BibitemShut {NoStop}%
\bibitem [{\citenamefont {Lv}\ \emph {et~al.}(2015{\natexlab{a}})\citenamefont
  {Lv}, \citenamefont {Weng}, \citenamefont {Fu}, \citenamefont {Wang},
  \citenamefont {Miao}, \citenamefont {Ma}, \citenamefont {Richard},
  \citenamefont {Huang}, \citenamefont {Zhao}, \citenamefont {Chen},
  \citenamefont {Fang}, \citenamefont {Dai}, \citenamefont {Qian},\ and\
  \citenamefont {Ding}}]{lv2015}%
  \BibitemOpen
  \bibfield  {author} {\bibinfo {author} {\bibfnamefont {B.~Q.}\ \bibnamefont
  {Lv}}, \bibinfo {author} {\bibfnamefont {H.~M.}\ \bibnamefont {Weng}},
  \bibinfo {author} {\bibfnamefont {B.~B.}\ \bibnamefont {Fu}}, \bibinfo
  {author} {\bibfnamefont {X.~P.}\ \bibnamefont {Wang}}, \bibinfo {author}
  {\bibfnamefont {H.}~\bibnamefont {Miao}}, \bibinfo {author} {\bibfnamefont
  {J.}~\bibnamefont {Ma}}, \bibinfo {author} {\bibfnamefont {P.}~\bibnamefont
  {Richard}}, \bibinfo {author} {\bibfnamefont {X.~C.}\ \bibnamefont {Huang}},
  \bibinfo {author} {\bibfnamefont {L.~X.}\ \bibnamefont {Zhao}}, \bibinfo
  {author} {\bibfnamefont {G.~F.}\ \bibnamefont {Chen}}, \bibinfo {author}
  {\bibfnamefont {Z.}~\bibnamefont {Fang}}, \bibinfo {author} {\bibfnamefont
  {X.}~\bibnamefont {Dai}}, \bibinfo {author} {\bibfnamefont {T.}~\bibnamefont
  {Qian}}, \ and\ \bibinfo {author} {\bibfnamefont {H.}~\bibnamefont {Ding}},\
  }\href {http://journals.aps.org/prx/abstract/10.1103/PhysRevX.5.031013}
  {\bibfield  {journal} {\bibinfo  {journal} {Phys. Rev. X}\ }\textbf {\bibinfo
  {volume} {5}},\ \bibinfo {pages} {031013} (\bibinfo {year}
  {2015}{\natexlab{a}})}\BibitemShut {NoStop}%
\bibitem [{\citenamefont {Xu}\ \emph {et~al.}(2015{\natexlab{a}})\citenamefont
  {Xu}, \citenamefont {Belopolski}, \citenamefont {Alidoust}, \citenamefont
  {Neupane}, \citenamefont {Bian}, \citenamefont {Zhang}, \citenamefont
  {Sankar}, \citenamefont {Chang}, \citenamefont {Yuan}, \citenamefont {Lee},
  \citenamefont {Huang}, \citenamefont {Zheng}, \citenamefont {Ma},
  \citenamefont {Sanchez}, \citenamefont {Wang}, \citenamefont {Bansil},
  \citenamefont {Chou}, \citenamefont {Shibayev}, \citenamefont {Lin},
  \citenamefont {Jia},\ and\ \citenamefont {Hasan}}]{xu2015}%
  \BibitemOpen
  \bibfield  {author} {\bibinfo {author} {\bibfnamefont {S.-Y.}\ \bibnamefont
  {Xu}}, \bibinfo {author} {\bibfnamefont {I.}~\bibnamefont {Belopolski}},
  \bibinfo {author} {\bibfnamefont {N.}~\bibnamefont {Alidoust}}, \bibinfo
  {author} {\bibfnamefont {M.}~\bibnamefont {Neupane}}, \bibinfo {author}
  {\bibfnamefont {G.}~\bibnamefont {Bian}}, \bibinfo {author} {\bibfnamefont
  {C.}~\bibnamefont {Zhang}}, \bibinfo {author} {\bibfnamefont
  {R.}~\bibnamefont {Sankar}}, \bibinfo {author} {\bibfnamefont
  {G.}~\bibnamefont {Chang}}, \bibinfo {author} {\bibfnamefont
  {Z.}~\bibnamefont {Yuan}}, \bibinfo {author} {\bibfnamefont {C.-C.}\
  \bibnamefont {Lee}}, \bibinfo {author} {\bibfnamefont {S.-M.}\ \bibnamefont
  {Huang}}, \bibinfo {author} {\bibfnamefont {H.}~\bibnamefont {Zheng}},
  \bibinfo {author} {\bibfnamefont {J.}~\bibnamefont {Ma}}, \bibinfo {author}
  {\bibfnamefont {D.~S.}\ \bibnamefont {Sanchez}}, \bibinfo {author}
  {\bibfnamefont {B.}~\bibnamefont {Wang}}, \bibinfo {author} {\bibfnamefont
  {A.}~\bibnamefont {Bansil}}, \bibinfo {author} {\bibfnamefont
  {F.}~\bibnamefont {Chou}}, \bibinfo {author} {\bibfnamefont {P.~P.}\
  \bibnamefont {Shibayev}}, \bibinfo {author} {\bibfnamefont {H.}~\bibnamefont
  {Lin}}, \bibinfo {author} {\bibfnamefont {S.}~\bibnamefont {Jia}}, \ and\
  \bibinfo {author} {\bibfnamefont {M.~Z.}\ \bibnamefont {Hasan}},\ }\href
  {http://www.sciencemag.org/content/349/6248/613.short} {\bibfield  {journal}
  {\bibinfo  {journal} {Science}\ }\textbf {\bibinfo {volume} {349}},\ \bibinfo
  {pages} {613} (\bibinfo {year} {2015}{\natexlab{a}})}\BibitemShut {NoStop}%
\bibitem [{\citenamefont {Yang}\ \emph {et~al.}(2015)\citenamefont {Yang},
  \citenamefont {Liu}, \citenamefont {Sun}, \citenamefont {Peng}, \citenamefont
  {Yang}, \citenamefont {Zhang}, \citenamefont {Zhou}, \citenamefont {Zhang},
  \citenamefont {Guo}, \citenamefont {Rahn}, \citenamefont {Prabhakaran},
  \citenamefont {Hussain}, \citenamefont {Mo}, \citenamefont {Felser},
  \citenamefont {Yan},\ and\ \citenamefont {Chen}}]{yang2015}%
  \BibitemOpen
  \bibfield  {author} {\bibinfo {author} {\bibfnamefont {L.~X.}\ \bibnamefont
  {Yang}}, \bibinfo {author} {\bibfnamefont {Z.~K.}\ \bibnamefont {Liu}},
  \bibinfo {author} {\bibfnamefont {Y.}~\bibnamefont {Sun}}, \bibinfo {author}
  {\bibfnamefont {H.}~\bibnamefont {Peng}}, \bibinfo {author} {\bibfnamefont
  {H.~F.}\ \bibnamefont {Yang}}, \bibinfo {author} {\bibfnamefont
  {T.}~\bibnamefont {Zhang}}, \bibinfo {author} {\bibfnamefont
  {B.}~\bibnamefont {Zhou}}, \bibinfo {author} {\bibfnamefont {Y.}~\bibnamefont
  {Zhang}}, \bibinfo {author} {\bibfnamefont {Y.~F.}\ \bibnamefont {Guo}},
  \bibinfo {author} {\bibfnamefont {M.}~\bibnamefont {Rahn}}, \bibinfo {author}
  {\bibfnamefont {D.}~\bibnamefont {Prabhakaran}}, \bibinfo {author}
  {\bibfnamefont {Z.}~\bibnamefont {Hussain}}, \bibinfo {author} {\bibfnamefont
  {S.~K.}\ \bibnamefont {Mo}}, \bibinfo {author} {\bibfnamefont
  {C.}~\bibnamefont {Felser}}, \bibinfo {author} {\bibfnamefont
  {B.}~\bibnamefont {Yan}}, \ and\ \bibinfo {author} {\bibfnamefont {Y.~L.}\
  \bibnamefont {Chen}},\ }\href {\doibase 10.1038/nphys3425} {\bibfield
  {journal} {\bibinfo  {journal} {Nat. Phys.}\ }\textbf {\bibinfo {volume}
  {11}},\ \bibinfo {pages} {728} (\bibinfo {year} {2015})}\BibitemShut
  {NoStop}%
\bibitem [{\citenamefont {Xu}\ \emph {et~al.}(2015{\natexlab{b}})\citenamefont
  {Xu}, \citenamefont {Alidoust}, \citenamefont {Belopolski}, \citenamefont
  {Yuan}, \citenamefont {Bian}, \citenamefont {Chang}, \citenamefont {Zheng},
  \citenamefont {Strocov}, \citenamefont {Sanchez}, \citenamefont {Chang},
  \citenamefont {Zhang}, \citenamefont {Mou}, \citenamefont {Wu}, \citenamefont
  {Huang}, \citenamefont {Lee}, \citenamefont {Huang}, \citenamefont {Wang},
  \citenamefont {Bansil}, \citenamefont {Jeng}, \citenamefont {Neupert},
  \citenamefont {Kaminski}, \citenamefont {Lin}, \citenamefont {Jia},\ and\
  \citenamefont {Zahid~Hasan}}]{xu2015a}%
  \BibitemOpen
  \bibfield  {author} {\bibinfo {author} {\bibfnamefont {S.-Y.}\ \bibnamefont
  {Xu}}, \bibinfo {author} {\bibfnamefont {N.}~\bibnamefont {Alidoust}},
  \bibinfo {author} {\bibfnamefont {I.}~\bibnamefont {Belopolski}}, \bibinfo
  {author} {\bibfnamefont {Z.}~\bibnamefont {Yuan}}, \bibinfo {author}
  {\bibfnamefont {G.}~\bibnamefont {Bian}}, \bibinfo {author} {\bibfnamefont
  {T.-R.}\ \bibnamefont {Chang}}, \bibinfo {author} {\bibfnamefont
  {H.}~\bibnamefont {Zheng}}, \bibinfo {author} {\bibfnamefont {V.~N.}\
  \bibnamefont {Strocov}}, \bibinfo {author} {\bibfnamefont {D.~S.}\
  \bibnamefont {Sanchez}}, \bibinfo {author} {\bibfnamefont {G.}~\bibnamefont
  {Chang}}, \bibinfo {author} {\bibfnamefont {C.}~\bibnamefont {Zhang}},
  \bibinfo {author} {\bibfnamefont {D.}~\bibnamefont {Mou}}, \bibinfo {author}
  {\bibfnamefont {Y.}~\bibnamefont {Wu}}, \bibinfo {author} {\bibfnamefont
  {L.}~\bibnamefont {Huang}}, \bibinfo {author} {\bibfnamefont {C.-C.}\
  \bibnamefont {Lee}}, \bibinfo {author} {\bibfnamefont {S.-M.}\ \bibnamefont
  {Huang}}, \bibinfo {author} {\bibfnamefont {B.}~\bibnamefont {Wang}},
  \bibinfo {author} {\bibfnamefont {A.}~\bibnamefont {Bansil}}, \bibinfo
  {author} {\bibfnamefont {H.-T.}\ \bibnamefont {Jeng}}, \bibinfo {author}
  {\bibfnamefont {T.}~\bibnamefont {Neupert}}, \bibinfo {author} {\bibfnamefont
  {A.}~\bibnamefont {Kaminski}}, \bibinfo {author} {\bibfnamefont
  {H.}~\bibnamefont {Lin}}, \bibinfo {author} {\bibfnamefont {S.}~\bibnamefont
  {Jia}}, \ and\ \bibinfo {author} {\bibfnamefont {M.}~\bibnamefont
  {Zahid~Hasan}},\ }\href {\doibase 10.1038/nphys3437} {\bibfield  {journal}
  {\bibinfo  {journal} {Nat. Phys.}\ }\textbf {\bibinfo {volume} {11}},\
  \bibinfo {pages} {748} (\bibinfo {year} {2015}{\natexlab{b}})}\BibitemShut
  {NoStop}%
\bibitem [{\citenamefont {Lv}\ \emph {et~al.}(2015{\natexlab{b}})\citenamefont
  {Lv}, \citenamefont {Xu}, \citenamefont {Weng}, \citenamefont {Ma},
  \citenamefont {Richard}, \citenamefont {Huang}, \citenamefont {Zhao},
  \citenamefont {Chen}, \citenamefont {Matt}, \citenamefont {Bisti},
  \citenamefont {Strocov}, \citenamefont {Mesot}, \citenamefont {Fang},
  \citenamefont {Dai}, \citenamefont {Qian}, \citenamefont {Shi},\ and\
  \citenamefont {Ding}}]{lv2015a}%
  \BibitemOpen
  \bibfield  {author} {\bibinfo {author} {\bibfnamefont {B.~Q.}\ \bibnamefont
  {Lv}}, \bibinfo {author} {\bibfnamefont {N.}~\bibnamefont {Xu}}, \bibinfo
  {author} {\bibfnamefont {H.~M.}\ \bibnamefont {Weng}}, \bibinfo {author}
  {\bibfnamefont {J.~Z.}\ \bibnamefont {Ma}}, \bibinfo {author} {\bibfnamefont
  {P.}~\bibnamefont {Richard}}, \bibinfo {author} {\bibfnamefont {X.~C.}\
  \bibnamefont {Huang}}, \bibinfo {author} {\bibfnamefont {L.~X.}\ \bibnamefont
  {Zhao}}, \bibinfo {author} {\bibfnamefont {G.~F.}\ \bibnamefont {Chen}},
  \bibinfo {author} {\bibfnamefont {C.~E.}\ \bibnamefont {Matt}}, \bibinfo
  {author} {\bibfnamefont {F.}~\bibnamefont {Bisti}}, \bibinfo {author}
  {\bibfnamefont {V.~N.}\ \bibnamefont {Strocov}}, \bibinfo {author}
  {\bibfnamefont {J.}~\bibnamefont {Mesot}}, \bibinfo {author} {\bibfnamefont
  {Z.}~\bibnamefont {Fang}}, \bibinfo {author} {\bibfnamefont {X.}~\bibnamefont
  {Dai}}, \bibinfo {author} {\bibfnamefont {T.}~\bibnamefont {Qian}}, \bibinfo
  {author} {\bibfnamefont {M.}~\bibnamefont {Shi}}, \ and\ \bibinfo {author}
  {\bibfnamefont {H.}~\bibnamefont {Ding}},\ }\href {\doibase
  10.1038/nphys3426} {\bibfield  {journal} {\bibinfo  {journal} {Nat. Phys.}\
  }\textbf {\bibinfo {volume} {11}},\ \bibinfo {pages} {724} (\bibinfo {year}
  {2015}{\natexlab{b}})}\BibitemShut {NoStop}%
\bibitem [{\citenamefont {Kim}\ \emph {et~al.}(2013)\citenamefont {Kim},
  \citenamefont {Kim}, \citenamefont {Wang}, \citenamefont {Sasaki},
  \citenamefont {Satoh}, \citenamefont {Ohnishi}, \citenamefont {Kitaura},
  \citenamefont {Yang},\ and\ \citenamefont {Li}}]{Kim13prl}%
  \BibitemOpen
  \bibfield  {author} {\bibinfo {author} {\bibfnamefont {H.~J.}\ \bibnamefont
  {Kim}}, \bibinfo {author} {\bibfnamefont {K.~S.}\ \bibnamefont {Kim}},
  \bibinfo {author} {\bibfnamefont {J.~F.}\ \bibnamefont {Wang}}, \bibinfo
  {author} {\bibfnamefont {M.}~\bibnamefont {Sasaki}}, \bibinfo {author}
  {\bibfnamefont {N.}~\bibnamefont {Satoh}}, \bibinfo {author} {\bibfnamefont
  {A.}~\bibnamefont {Ohnishi}}, \bibinfo {author} {\bibfnamefont
  {M.}~\bibnamefont {Kitaura}}, \bibinfo {author} {\bibfnamefont
  {M.}~\bibnamefont {Yang}}, \ and\ \bibinfo {author} {\bibfnamefont
  {L.}~\bibnamefont {Li}},\ }\href {\doibase 10.1103/PhysRevLett.111.246603}
  {\bibfield  {journal} {\bibinfo  {journal} {Phys. Rev. Lett.}\ }\textbf
  {\bibinfo {volume} {111}},\ \bibinfo {pages} {246603} (\bibinfo {year}
  {2013})}\BibitemShut {NoStop}%
\bibitem [{\citenamefont {Li}\ \emph {et~al.}(2016{\natexlab{a}})\citenamefont
  {Li}, \citenamefont {Kharzeev}, \citenamefont {Zhang}, \citenamefont {Huang},
  \citenamefont {Pletikosi{\'c}}, \citenamefont {Fedorov}, \citenamefont
  {Zhong}, \citenamefont {Schneeloch}, \citenamefont {Gu},\ and\ \citenamefont
  {Valla}}]{Li14arXiv}%
  \BibitemOpen
  \bibfield  {author} {\bibinfo {author} {\bibfnamefont {Q.}~\bibnamefont
  {Li}}, \bibinfo {author} {\bibfnamefont {D.~E.}\ \bibnamefont {Kharzeev}},
  \bibinfo {author} {\bibfnamefont {C.}~\bibnamefont {Zhang}}, \bibinfo
  {author} {\bibfnamefont {Y.}~\bibnamefont {Huang}}, \bibinfo {author}
  {\bibfnamefont {I.}~\bibnamefont {Pletikosi{\'c}}}, \bibinfo {author}
  {\bibfnamefont {A.}~\bibnamefont {Fedorov}}, \bibinfo {author} {\bibfnamefont
  {R.}~\bibnamefont {Zhong}}, \bibinfo {author} {\bibfnamefont
  {J.}~\bibnamefont {Schneeloch}}, \bibinfo {author} {\bibfnamefont
  {G.}~\bibnamefont {Gu}}, \ and\ \bibinfo {author} {\bibfnamefont
  {T.}~\bibnamefont {Valla}},\ }\href
  {http://www.nature.com/nphys/journal/vaop/ncurrent/full/nphys3648.html}
  {\bibfield  {journal} {\bibinfo  {journal} {Nat. Phys.}\ } (\bibinfo {year}
  {2016}{\natexlab{a}})}\BibitemShut {NoStop}%
\bibitem [{\citenamefont {Huang}\ \emph
  {et~al.}(2015{\natexlab{b}})\citenamefont {Huang}, \citenamefont {Zhao},
  \citenamefont {Long}, \citenamefont {Wang}, \citenamefont {Chen},
  \citenamefont {Yang}, \citenamefont {Liang}, \citenamefont {Xue},
  \citenamefont {Weng}, \citenamefont {Fang}, \citenamefont {Dai},\ and\
  \citenamefont {Chen}}]{HuangXC15prx}%
  \BibitemOpen
  \bibfield  {author} {\bibinfo {author} {\bibfnamefont {X.~C.}\ \bibnamefont
  {Huang}}, \bibinfo {author} {\bibfnamefont {L.~X.}\ \bibnamefont {Zhao}},
  \bibinfo {author} {\bibfnamefont {Y.~J.}\ \bibnamefont {Long}}, \bibinfo
  {author} {\bibfnamefont {P.~P.}\ \bibnamefont {Wang}}, \bibinfo {author}
  {\bibfnamefont {D.}~\bibnamefont {Chen}}, \bibinfo {author} {\bibfnamefont
  {Z.~H.}\ \bibnamefont {Yang}}, \bibinfo {author} {\bibfnamefont
  {H.}~\bibnamefont {Liang}}, \bibinfo {author} {\bibfnamefont {M.~Q.}\
  \bibnamefont {Xue}}, \bibinfo {author} {\bibfnamefont {H.~M.}\ \bibnamefont
  {Weng}}, \bibinfo {author} {\bibfnamefont {Z.}~\bibnamefont {Fang}}, \bibinfo
  {author} {\bibfnamefont {X.}~\bibnamefont {Dai}}, \ and\ \bibinfo {author}
  {\bibfnamefont {G.~F.}\ \bibnamefont {Chen}},\ }\href {\doibase
  10.1103/PhysRevX.5.031023} {\bibfield  {journal} {\bibinfo  {journal} {Phys.
  Rev. X}\ }\textbf {\bibinfo {volume} {5}},\ \bibinfo {pages} {031023}
  (\bibinfo {year} {2015}{\natexlab{b}})}\BibitemShut {NoStop}%
\bibitem [{\citenamefont {Zhang}\ \emph {et~al.}(2016)\citenamefont {Zhang},
  \citenamefont {Xu}, \citenamefont {Belopolski}, \citenamefont {Yuan},
  \citenamefont {Lin}, \citenamefont {Tong}, \citenamefont {Alidoust},
  \citenamefont {Lee}, \citenamefont {Huang}, \citenamefont {Lin},
  \citenamefont {Neupane}, \citenamefont {Sanchez}, \citenamefont {Zheng},
  \citenamefont {Bian}, \citenamefont {Wang}, \citenamefont {Zhang},
  \citenamefont {Lu}, \citenamefont {Shen}, \citenamefont {Neupert},
  \citenamefont {Hasan},\ and\ \citenamefont {Jia}}]{ZhangCL15arXiv}%
  \BibitemOpen
  \bibfield  {author} {\bibinfo {author} {\bibfnamefont {C.}~\bibnamefont
  {Zhang}}, \bibinfo {author} {\bibfnamefont {S.~Y.}\ \bibnamefont {Xu}},
  \bibinfo {author} {\bibfnamefont {I.}~\bibnamefont {Belopolski}}, \bibinfo
  {author} {\bibfnamefont {Z.}~\bibnamefont {Yuan}}, \bibinfo {author}
  {\bibfnamefont {Z.}~\bibnamefont {Lin}}, \bibinfo {author} {\bibfnamefont
  {B.}~\bibnamefont {Tong}}, \bibinfo {author} {\bibfnamefont {N.}~\bibnamefont
  {Alidoust}}, \bibinfo {author} {\bibfnamefont {C.~C.}\ \bibnamefont {Lee}},
  \bibinfo {author} {\bibfnamefont {S.~M.}\ \bibnamefont {Huang}}, \bibinfo
  {author} {\bibfnamefont {H.}~\bibnamefont {Lin}}, \bibinfo {author}
  {\bibfnamefont {M.}~\bibnamefont {Neupane}}, \bibinfo {author} {\bibfnamefont
  {D.~S.}\ \bibnamefont {Sanchez}}, \bibinfo {author} {\bibfnamefont
  {H.}~\bibnamefont {Zheng}}, \bibinfo {author} {\bibfnamefont
  {G.}~\bibnamefont {Bian}}, \bibinfo {author} {\bibfnamefont {J.}~\bibnamefont
  {Wang}}, \bibinfo {author} {\bibfnamefont {C.}~\bibnamefont {Zhang}},
  \bibinfo {author} {\bibfnamefont {H.~Z.}\ \bibnamefont {Lu}}, \bibinfo
  {author} {\bibfnamefont {S.~Q.}\ \bibnamefont {Shen}}, \bibinfo {author}
  {\bibfnamefont {T.}~\bibnamefont {Neupert}}, \bibinfo {author} {\bibfnamefont
  {M.~Z.}\ \bibnamefont {Hasan}}, \ and\ \bibinfo {author} {\bibfnamefont
  {S.}~\bibnamefont {Jia}},\ }\href
  {http://www.nature.com/ncomms/2016/160225/ncomms10735/full/ncomms10735.html}
  {\bibfield  {journal} {\bibinfo  {journal} {Nat. Commun.}\ }\textbf {\bibinfo
  {volume} {7}},\ \bibinfo {pages} {10735} (\bibinfo {year}
  {2016})}\BibitemShut {NoStop}%
\bibitem [{\citenamefont {Shekhar}\ \emph {et~al.}(2015)\citenamefont
  {Shekhar}, \citenamefont {Nayak}, \citenamefont {Sun}, \citenamefont
  {Schmidt}, \citenamefont {Nicklas}, \citenamefont {Leermakers}, \citenamefont
  {Zeitler}, \citenamefont {Skourski}, \citenamefont {Wosnitza}, \citenamefont
  {Liu}, \citenamefont {Chen}, \citenamefont {Schnelle}, \citenamefont
  {Borrmann}, \citenamefont {Grin}, \citenamefont {Felser},\ and\ \citenamefont
  {Yan}}]{Binghai15}%
  \BibitemOpen
  \bibfield  {author} {\bibinfo {author} {\bibfnamefont {C.}~\bibnamefont
  {Shekhar}}, \bibinfo {author} {\bibfnamefont {A.~K.}\ \bibnamefont {Nayak}},
  \bibinfo {author} {\bibfnamefont {Y.}~\bibnamefont {Sun}}, \bibinfo {author}
  {\bibfnamefont {M.}~\bibnamefont {Schmidt}}, \bibinfo {author} {\bibfnamefont
  {M.}~\bibnamefont {Nicklas}}, \bibinfo {author} {\bibfnamefont
  {I.}~\bibnamefont {Leermakers}}, \bibinfo {author} {\bibfnamefont
  {U.}~\bibnamefont {Zeitler}}, \bibinfo {author} {\bibfnamefont
  {Y.}~\bibnamefont {Skourski}}, \bibinfo {author} {\bibfnamefont
  {J.}~\bibnamefont {Wosnitza}}, \bibinfo {author} {\bibfnamefont
  {Z.}~\bibnamefont {Liu}}, \bibinfo {author} {\bibfnamefont {Y.}~\bibnamefont
  {Chen}}, \bibinfo {author} {\bibfnamefont {W.}~\bibnamefont {Schnelle}},
  \bibinfo {author} {\bibfnamefont {H.}~\bibnamefont {Borrmann}}, \bibinfo
  {author} {\bibfnamefont {Y.}~\bibnamefont {Grin}}, \bibinfo {author}
  {\bibfnamefont {C.}~\bibnamefont {Felser}}, \ and\ \bibinfo {author}
  {\bibfnamefont {B.}~\bibnamefont {Yan}},\ }\href
  {http://www.nature.com/nphys/journal/v11/n8/full/nphys3372.html} {\bibfield
  {journal} {\bibinfo  {journal} {Nat. Phys.}\ }\textbf {\bibinfo {volume}
  {11}},\ \bibinfo {pages} {645} (\bibinfo {year} {2015})}\BibitemShut
  {NoStop}%
\bibitem [{\citenamefont {Xiong}\ \emph {et~al.}(2015)\citenamefont {Xiong},
  \citenamefont {Kushwaha}, \citenamefont {Liang}, \citenamefont {Krizan},
  \citenamefont {Hirschberger}, \citenamefont {Wang}, \citenamefont {Cava},\
  and\ \citenamefont {Ong}}]{ong2015}%
  \BibitemOpen
  \bibfield  {author} {\bibinfo {author} {\bibfnamefont {J.}~\bibnamefont
  {Xiong}}, \bibinfo {author} {\bibfnamefont {S.~K.}\ \bibnamefont {Kushwaha}},
  \bibinfo {author} {\bibfnamefont {T.}~\bibnamefont {Liang}}, \bibinfo
  {author} {\bibfnamefont {J.~W.}\ \bibnamefont {Krizan}}, \bibinfo {author}
  {\bibfnamefont {M.}~\bibnamefont {Hirschberger}}, \bibinfo {author}
  {\bibfnamefont {W.}~\bibnamefont {Wang}}, \bibinfo {author} {\bibfnamefont
  {R.}~\bibnamefont {Cava}}, \ and\ \bibinfo {author} {\bibfnamefont
  {N.}~\bibnamefont {Ong}},\ }\href
  {http://www.sciencemag.org/content/350/6259/413.short} {\bibfield  {journal}
  {\bibinfo  {journal} {Science}\ }\textbf {\bibinfo {volume} {350}},\ \bibinfo
  {pages} {413} (\bibinfo {year} {2015})}\BibitemShut {NoStop}%
\bibitem [{\citenamefont {Wang}\ \emph {et~al.}(2016)\citenamefont {Wang},
  \citenamefont {Zheng}, \citenamefont {Shen}, \citenamefont {Lu},
  \citenamefont {Fang}, \citenamefont {Sheng}, \citenamefont {Zhou},
  \citenamefont {Yang}, \citenamefont {Li}, \citenamefont {Feng},\ and\
  \citenamefont {Xu}}]{zhouyi15}%
  \BibitemOpen
  \bibfield  {author} {\bibinfo {author} {\bibfnamefont {Z.}~\bibnamefont
  {Wang}}, \bibinfo {author} {\bibfnamefont {Y.}~\bibnamefont {Zheng}},
  \bibinfo {author} {\bibfnamefont {Z.}~\bibnamefont {Shen}}, \bibinfo {author}
  {\bibfnamefont {Y.}~\bibnamefont {Lu}}, \bibinfo {author} {\bibfnamefont
  {H.}~\bibnamefont {Fang}}, \bibinfo {author} {\bibfnamefont {F.}~\bibnamefont
  {Sheng}}, \bibinfo {author} {\bibfnamefont {Y.}~\bibnamefont {Zhou}},
  \bibinfo {author} {\bibfnamefont {X.}~\bibnamefont {Yang}}, \bibinfo {author}
  {\bibfnamefont {Y.}~\bibnamefont {Li}}, \bibinfo {author} {\bibfnamefont
  {C.}~\bibnamefont {Feng}}, \ and\ \bibinfo {author} {\bibfnamefont {Z.-A.}\
  \bibnamefont {Xu}},\ }\href {\doibase 10.1103/PhysRevB.93.121112} {\bibfield
  {journal} {\bibinfo  {journal} {Phys. Rev. B}\ }\textbf {\bibinfo {volume}
  {93}},\ \bibinfo {pages} {121112} (\bibinfo {year} {2016})}\BibitemShut
  {NoStop}%
\bibitem [{\citenamefont {Li}\ \emph {et~al.}(2015)\citenamefont {Li},
  \citenamefont {Wang}, \citenamefont {Liu}, \citenamefont {Wang},
  \citenamefont {Liao},\ and\ \citenamefont {Yu}}]{li2015nc}%
  \BibitemOpen
  \bibfield  {author} {\bibinfo {author} {\bibfnamefont {C.-Z.}\ \bibnamefont
  {Li}}, \bibinfo {author} {\bibfnamefont {L.-X.}\ \bibnamefont {Wang}},
  \bibinfo {author} {\bibfnamefont {H.}~\bibnamefont {Liu}}, \bibinfo {author}
  {\bibfnamefont {J.}~\bibnamefont {Wang}}, \bibinfo {author} {\bibfnamefont
  {Z.-M.}\ \bibnamefont {Liao}}, \ and\ \bibinfo {author} {\bibfnamefont
  {D.-P.}\ \bibnamefont {Yu}},\ }\href
  {http://www.nature.com/ncomms/2015/151217/ncomms10137/full/ncomms10137.html}
  {\bibfield  {journal} {\bibinfo  {journal} {Nat. Commun.}\ }\textbf {\bibinfo
  {volume} {6}} (\bibinfo {year} {2015})}\BibitemShut {NoStop}%
\bibitem [{\citenamefont {Li}\ \emph {et~al.}(2016{\natexlab{b}})\citenamefont
  {Li}, \citenamefont {He}, \citenamefont {Lu}, \citenamefont {Zhang},
  \citenamefont {Liu}, \citenamefont {Ma}, \citenamefont {Fan}, \citenamefont
  {Shen},\ and\ \citenamefont {Wang}}]{Li2016}%
  \BibitemOpen
  \bibfield  {author} {\bibinfo {author} {\bibfnamefont {H.}~\bibnamefont
  {Li}}, \bibinfo {author} {\bibfnamefont {H.}~\bibnamefont {He}}, \bibinfo
  {author} {\bibfnamefont {H.-Z.}\ \bibnamefont {Lu}}, \bibinfo {author}
  {\bibfnamefont {H.}~\bibnamefont {Zhang}}, \bibinfo {author} {\bibfnamefont
  {H.}~\bibnamefont {Liu}}, \bibinfo {author} {\bibfnamefont {R.}~\bibnamefont
  {Ma}}, \bibinfo {author} {\bibfnamefont {Z.}~\bibnamefont {Fan}}, \bibinfo
  {author} {\bibfnamefont {S.-Q.}\ \bibnamefont {Shen}}, \ and\ \bibinfo
  {author} {\bibfnamefont {J.}~\bibnamefont {Wang}},\ }\href
  {http://www.nature.com/ncomms/2016/160108/ncomms10301/full/ncomms10301.html}
  {\bibfield  {journal} {\bibinfo  {journal} {Nat. Commun.}\ }\textbf {\bibinfo
  {volume} {7}} (\bibinfo {year} {2016}{\natexlab{b}})}\BibitemShut {NoStop}%
\bibitem [{\citenamefont {Lu}\ and\ \citenamefont
  {Shen}(2015)}]{Lu15Weyl-Localization}%
  \BibitemOpen
  \bibfield  {author} {\bibinfo {author} {\bibfnamefont {H.~Z.}\ \bibnamefont
  {Lu}}\ and\ \bibinfo {author} {\bibfnamefont {S.~Q.}\ \bibnamefont {Shen}},\
  }\href {\doibase 10.1103/PhysRevB.92.035203} {\bibfield  {journal} {\bibinfo
  {journal} {Phys. Rev. B}\ }\textbf {\bibinfo {volume} {92}},\ \bibinfo
  {pages} {035203} (\bibinfo {year} {2015})}\BibitemShut {NoStop}%
\bibitem [{\citenamefont {Xiao}\ \emph {et~al.}(2010)\citenamefont {Xiao},
  \citenamefont {Chang},\ and\ \citenamefont {Niu}}]{Xiao10rmp}%
  \BibitemOpen
  \bibfield  {author} {\bibinfo {author} {\bibfnamefont {D.}~\bibnamefont
  {Xiao}}, \bibinfo {author} {\bibfnamefont {M.~C.}\ \bibnamefont {Chang}}, \
  and\ \bibinfo {author} {\bibfnamefont {Q.}~\bibnamefont {Niu}},\ }\href
  {\doibase 10.1103/RevModPhys.82.1959} {\bibfield  {journal} {\bibinfo
  {journal} {Rev. Mod. Phys.}\ }\textbf {\bibinfo {volume} {82}},\ \bibinfo
  {pages} {1959} (\bibinfo {year} {2010})}\BibitemShut {NoStop}%
\bibitem [{\citenamefont {Lee}\ and\ \citenamefont
  {Ramakrishnan}(1985)}]{Lee85rmp}%
  \BibitemOpen
  \bibfield  {author} {\bibinfo {author} {\bibfnamefont {P.~A.}\ \bibnamefont
  {Lee}}\ and\ \bibinfo {author} {\bibfnamefont {T.~V.}\ \bibnamefont
  {Ramakrishnan}},\ }\href {\doibase 10.1103/RevModPhys.57.287} {\bibfield
  {journal} {\bibinfo  {journal} {Rev. Mod. Phys.}\ }\textbf {\bibinfo {volume}
  {57}},\ \bibinfo {pages} {287} (\bibinfo {year} {1985})}\BibitemShut
  {NoStop}%
\bibitem [{Sup()}]{Supp-Weyl}%
  \BibitemOpen
  \href@noop {} {\bibinfo  {journal} {See Supplemental Materials for the
  detailed calculations}\ }\BibitemShut {NoStop}%
\bibitem [{\citenamefont {Altland}\ and\ \citenamefont
  {Zirnbauer}(1997)}]{Altland97prb}%
  \BibitemOpen
\bibfield  {journal} {  }\bibfield  {author} {\bibinfo {author} {\bibfnamefont
  {A.}~\bibnamefont {Altland}}\ and\ \bibinfo {author} {\bibfnamefont {M.~R.}\
  \bibnamefont {Zirnbauer}},\ }\href {\doibase 10.1103/PhysRevB.55.1142}
  {\bibfield  {journal} {\bibinfo  {journal} {Phys. Rev. B}\ }\textbf {\bibinfo
  {volume} {55}},\ \bibinfo {pages} {1142} (\bibinfo {year}
  {1997})}\BibitemShut {NoStop}%
\bibitem [{\citenamefont {Hikami}\ \emph {et~al.}(1980)\citenamefont {Hikami},
  \citenamefont {Larkin},\ and\ \citenamefont {Nagaoka}}]{Hikami80ptp}%
  \BibitemOpen
  \bibfield  {author} {\bibinfo {author} {\bibfnamefont {S.}~\bibnamefont
  {Hikami}}, \bibinfo {author} {\bibfnamefont {A.~I.}\ \bibnamefont {Larkin}},
  \ and\ \bibinfo {author} {\bibfnamefont {Y.}~\bibnamefont {Nagaoka}},\ }\href
  {\doibase 10.1143/PTP.63.707} {\bibfield  {journal} {\bibinfo  {journal}
  {Progr. Theor. Phys.}\ }\textbf {\bibinfo {volume} {63}},\ \bibinfo {pages}
  {707} (\bibinfo {year} {1980})}\BibitemShut {NoStop}%
\bibitem [{\citenamefont {Shon}\ and\ \citenamefont {Ando}(1998)}]{Shon98jpsj}%
  \BibitemOpen
  \bibfield  {author} {\bibinfo {author} {\bibfnamefont {N.~H.}\ \bibnamefont
  {Shon}}\ and\ \bibinfo {author} {\bibfnamefont {T.}~\bibnamefont {Ando}},\
  }\href {\doibase 10.1143/JPSJ.67.2421} {\bibfield  {journal} {\bibinfo
  {journal} {J. Phys. Soc. Jpn.}\ }\textbf {\bibinfo {volume} {67}},\ \bibinfo
  {pages} {2421} (\bibinfo {year} {1998})}\BibitemShut {NoStop}%
\bibitem [{\citenamefont {McCann}\ \emph {et~al.}(2006)\citenamefont {McCann},
  \citenamefont {Kechedzhi}, \citenamefont {Fal'ko}, \citenamefont {Suzuura},
  \citenamefont {Ando},\ and\ \citenamefont {Altshuler}}]{McCann06prl}%
  \BibitemOpen
  \bibfield  {author} {\bibinfo {author} {\bibfnamefont {E.}~\bibnamefont
  {McCann}}, \bibinfo {author} {\bibfnamefont {K.}~\bibnamefont {Kechedzhi}},
  \bibinfo {author} {\bibfnamefont {V.~I.}\ \bibnamefont {Fal'ko}}, \bibinfo
  {author} {\bibfnamefont {H.}~\bibnamefont {Suzuura}}, \bibinfo {author}
  {\bibfnamefont {T.}~\bibnamefont {Ando}}, \ and\ \bibinfo {author}
  {\bibfnamefont {B.~L.}\ \bibnamefont {Altshuler}},\ }\href {\doibase
  10.1103/PhysRevLett.97.146805} {\bibfield  {journal} {\bibinfo  {journal}
  {Phys. Rev. Lett.}\ }\textbf {\bibinfo {volume} {97}},\ \bibinfo {pages}
  {146805} (\bibinfo {year} {2006})}\BibitemShut {NoStop}%
\bibitem [{Note1()}]{Note1}%
  \BibitemOpen
  \bibinfo {note} {We only give the result for isotropic single-Weyl semimetals
  with $v_F=v_z=v_{\delimiter "026B30D }$; this simplification does not change
  any qualitative results with respect to quantum interference
  correction.}\BibitemShut {Stop}%
\bibitem [{Note2()}]{Note2}%
  \BibitemOpen
  \bibinfo {note} {We have carried out a coordinate transformation in deriving
  these results, where ${\mathchar '26\mkern -9muh}k_x=\protect \sqrt {k
  \protect \qopname \relax o{sin}\theta }\protect \qopname \relax o{cos}\varphi
  , {\mathchar '26\mkern -9muh}k_y=\protect \sqrt {k \protect \qopname \relax
  o{sin}\theta }\protect \qopname \relax o{sin}\varphi , 2{\mathchar '26\mkern
  -9muh}m v k_z =k \protect \qopname \relax o{cos}\theta $, $-{\protect \bf k}$
  is obtained by setting $\varphi \to \varphi + \pi $ and $\theta \to \pi
  -\theta $.}\BibitemShut {Stop}%
\bibitem [{\citenamefont {Son}\ and\ \citenamefont {Spivak}(2013)}]{Son13prb}%
  \BibitemOpen
  \bibfield  {author} {\bibinfo {author} {\bibfnamefont {D.~T.}\ \bibnamefont
  {Son}}\ and\ \bibinfo {author} {\bibfnamefont {B.~Z.}\ \bibnamefont
  {Spivak}},\ }\href {\doibase 10.1103/PhysRevB.88.104412} {\bibfield
  {journal} {\bibinfo  {journal} {Phys. Rev. B}\ }\textbf {\bibinfo {volume}
  {88}},\ \bibinfo {pages} {104412} (\bibinfo {year} {2013})}\BibitemShut
  {NoStop}%
\bibitem [{\citenamefont {Burkov}(2014)}]{Burkov14prl-chiral}%
  \BibitemOpen
  \bibfield  {author} {\bibinfo {author} {\bibfnamefont {A.~A.}\ \bibnamefont
  {Burkov}},\ }\href {\doibase 10.1103/PhysRevLett.113.247203} {\bibfield
  {journal} {\bibinfo  {journal} {Phys. Rev. Lett.}\ }\textbf {\bibinfo
  {volume} {113}},\ \bibinfo {pages} {247203} (\bibinfo {year}
  {2014})}\BibitemShut {NoStop}%
\bibitem [{\citenamefont {Zhang}\ \emph {et~al.}(2015)\citenamefont {Zhang},
  \citenamefont {Lu},\ and\ \citenamefont {Shen}}]{ZhangSB15arXiv}%
  \BibitemOpen
  \bibfield  {author} {\bibinfo {author} {\bibfnamefont {S.~B.}\ \bibnamefont
  {Zhang}}, \bibinfo {author} {\bibfnamefont {H.~Z.}\ \bibnamefont {Lu}}, \
  and\ \bibinfo {author} {\bibfnamefont {S.~Q.}\ \bibnamefont {Shen}},\ }\href
  {http://arxiv.org/abs/1509.02001} {\bibfield  {journal} {\bibinfo  {journal}
  {arXiv:1509.02001}\ } (\bibinfo {year} {2015})}\BibitemShut {NoStop}%
\bibitem [{\citenamefont {Suzuura}\ and\ \citenamefont
  {Ando}(2002)}]{Suzuura02prl}%
  \BibitemOpen
  \bibfield  {author} {\bibinfo {author} {\bibfnamefont {H.}~\bibnamefont
  {Suzuura}}\ and\ \bibinfo {author} {\bibfnamefont {T.}~\bibnamefont {Ando}},\
  }\href {\doibase 10.1103/PhysRevLett.89.266603} {\bibfield  {journal}
  {\bibinfo  {journal} {Phys. Rev. Lett.}\ }\textbf {\bibinfo {volume} {89}},\
  \bibinfo {pages} {266603} (\bibinfo {year} {2002})}\BibitemShut {NoStop}%
\bibitem [{\citenamefont {Tikhonenko}\ \emph {et~al.}(2009)\citenamefont
  {Tikhonenko}, \citenamefont {Kozikov}, \citenamefont {Savchenko},\ and\
  \citenamefont {Gorbachev}}]{Tikhonenko09prl}%
  \BibitemOpen
  \bibfield  {author} {\bibinfo {author} {\bibfnamefont {F.~V.}\ \bibnamefont
  {Tikhonenko}}, \bibinfo {author} {\bibfnamefont {A.~A.}\ \bibnamefont
  {Kozikov}}, \bibinfo {author} {\bibfnamefont {A.~K.}\ \bibnamefont
  {Savchenko}}, \ and\ \bibinfo {author} {\bibfnamefont {R.~V.}\ \bibnamefont
  {Gorbachev}},\ }\href {\doibase 10.1103/PhysRevLett.103.226801} {\bibfield
  {journal} {\bibinfo  {journal} {Phys. Rev. Lett.}\ }\textbf {\bibinfo
  {volume} {103}},\ \bibinfo {pages} {226801} (\bibinfo {year}
  {2009})}\BibitemShut {NoStop}%
\bibitem [{\citenamefont {Ando}\ \emph {et~al.}(1998)\citenamefont {Ando},
  \citenamefont {Nakanishi},\ and\ \citenamefont {Saito}}]{Ando98jpsj}%
  \BibitemOpen
  \bibfield  {author} {\bibinfo {author} {\bibfnamefont {T.}~\bibnamefont
  {Ando}}, \bibinfo {author} {\bibfnamefont {T.}~\bibnamefont {Nakanishi}}, \
  and\ \bibinfo {author} {\bibfnamefont {R.}~\bibnamefont {Saito}},\ }\href
  {\doibase 10.1143/JPSJ.67.2857} {\bibfield  {journal} {\bibinfo  {journal}
  {J. Phys. Soc. Jpn.}\ }\textbf {\bibinfo {volume} {67}},\ \bibinfo {pages}
  {2857} (\bibinfo {year} {1998})}\BibitemShut {NoStop}%
\bibitem [{\citenamefont {Lu}\ and\ \citenamefont {Shen}(2014)}]{Lu14prl}%
  \BibitemOpen
  \bibfield  {author} {\bibinfo {author} {\bibfnamefont {H.-Z.}\ \bibnamefont
  {Lu}}\ and\ \bibinfo {author} {\bibfnamefont {S.-Q.}\ \bibnamefont {Shen}},\
  }\href {\doibase 10.1103/PhysRevLett.112.146601} {\bibfield  {journal}
  {\bibinfo  {journal} {Phys. Rev. Lett.}\ }\textbf {\bibinfo {volume} {112}},\
  \bibinfo {pages} {146601} (\bibinfo {year} {2014})}\BibitemShut {NoStop}%
\end{thebibliography}
%

\appendix
\begin{widetext}
	
\section{SUPPLEMENTAL MATERIAL}
\renewcommand{\theequation}{S\arabic{equation}}
\setcounter{equation}{0}
\renewcommand{\thesection}{S\arabic{section}}
\renewcommand{\thetable}{S\arabic{table}}
\setcounter{table}{0}

\subsection{Appendix A: Berry phase and monopole charge}

The Hamiltonian can be rewritten as
\begin{equation}
\begin{split}
H
=E_{\bk}
\begin{pmatrix}
\chi \cos \theta & \sin \theta e^{i \mathcal{N} \varphi}\\
\sin \theta e^{-i \mathcal{N}\varphi} & -\chi \cos \theta
\end{pmatrix}
\equiv\bm{h} \cdot \bm{\sigma}
\end{split}
\end{equation}
where $k_{\pm}=k_x\pm i k_y$, $\cos \theta= v_z k_z/E_{\bk}$,
$\tan \varphi=k_y/k_x$, $k_{\Vert}^2=k_x^2+k_y^2$.
Taking the $\chi=+$ valley for example, the Berry curvature can be computed as \cite{Xiao10rmp}
\begin{equation}
\Omega _{\theta \varphi }=\partial _{\theta }A_{\varphi }-\partial _{\varphi }A_{\theta }=\frac{\mathcal{N}}{2}\sin \theta,
\end{equation}
where $A_{\theta }=\bra{u} i\partial_{\theta }u\rangle =0$, $A_{\varphi}=\bra{u}i\partial_{\varphi }u\rangle =\mathcal{N}\sin ^2\frac{\theta }{2}$ denotes the Berry connection and $u$ is the eigenstate.
Using
\begin{equation}
\Omega _{h_i,h_j}=\frac{1}{2}\frac{\partial (\varphi ,\cos  \theta )}{\partial \left(h_i,h_j\right)},
\end{equation}
we can rewrite the Berry curvature as a 3D vector via $\Omega _{\mu \nu }=\epsilon _{\mu \nu \xi }(\bm{\Omega })_{\xi }$ with $\epsilon _{\mu \nu \xi }$ the Levi-Civita anti-symmetric tensor. Then we obtain
\begin{equation}
\pmb{\Omega }=\frac{\mathcal{N}}{2}\frac{\pmb{h}}{h^3},
\end{equation}
which indicates that the magnitude of the Berry curvature on the Fermi sphere is uniform, independent of $\theta$ and $\varphi$.

Now we prove Eq. \eqref{charge_berry} for an arbitrary loop $\mathcal{C}$. To start, we consider the simplest case of $\mathcal{C}$ as a great circle in the equator defined by $\theta=\pi/2$ with a constant radius. We can define a specific orientation which encloses the northern hemisphere. The Berry phase along this great circle is calculated explicitly as
\begin{equation}\label{S4}
\gamma =\iint \bm{\Omega}\cdot \bm{n} \, dS=\int _0^{\pi /2}d\theta \int _0^{2\pi }d\varphi
\frac{\mathcal{N}}{2}\frac{\bm{h}}{h^3}\cdot \frac{\bm{h}}{h} h^2\sin  \theta=\mathcal{N}\pi.
\end{equation}

For the general case in which the loop $\mathcal{C}=P+\bar{P}$ is not a great circle, $\gamma$ is still quantized and amounts to half of the total flux, as proved below.
Consider a loop $\ell $ defined as
$\bk \to \bk_1 \to...\to \bk_n \to \bk$
where the orientation $\bn$ is defined by the right-hand rule and encloses an area $\mathcal{A}$.
An inversion operation maps $\ell$ to $\ell'$ that reads
$-\bk\to -\bk_1\to...\to -\bk_n \to -\bk$.
Note that this operation does not alter the orientation. If $\ell$, say, lies at the northern hemisphere,
$\ell'$ then lies at the southern hemisphere and hence encloses an area $\mathcal{A}'=4 \pi-\mathcal{A}$.
The loop we are interested in is defined as $\bk \to...\to -\bk \to...\to \bk$,
which is invariant under inversion, hence $\mathcal{A}=4 \pi-\mathcal{A}$,
i.e., the loop encloses half of the sphere. According to the Berry curvature given in Eq. \eqref{S4}, the Berry phase along the loop enclosing half of the sphere leading to an exact quantization, which proves Eq. \eqref{charge_berry}.

\subsection{Appendix B: Semi-classical Drude conductivity}
The detailed calculation for conductivities using Feynman diagram techniques in single-Weyl semimetals has been presented in Ref. \cite{Lu15Weyl-Localization};
therefore in this supplemental material we will mainly focus on double-Weyl semimetals.
Also for convenience we change the notation slightly with $v_z=v$ and $v_{\Vert}=1/2 m$.
To cope with the anisotropy inherited in double-Weyl semimetals, we perform a coordinate transformation which reads
\begin{equation}
\begin{split}
\hbar k_x =\sqrt{k \sin \theta }\cos \varphi,\quad \hbar k_y=\sqrt{k \sin \theta }\sin \varphi, \quad 2\hbar m v k_z= k \cos \theta,
\end{split}
\end{equation}
where $k>0$ and $\theta \in [0, \pi],\varphi \in [0, 2 \pi]$. 
The corresponding Jacobian is $k/4mv \hbar^3$.
After the transformation, $E_{\bk}=k/2m$, which is easier to handle.

\begin{figure}[ht]
\centering
\includegraphics[width=0.55\textwidth]{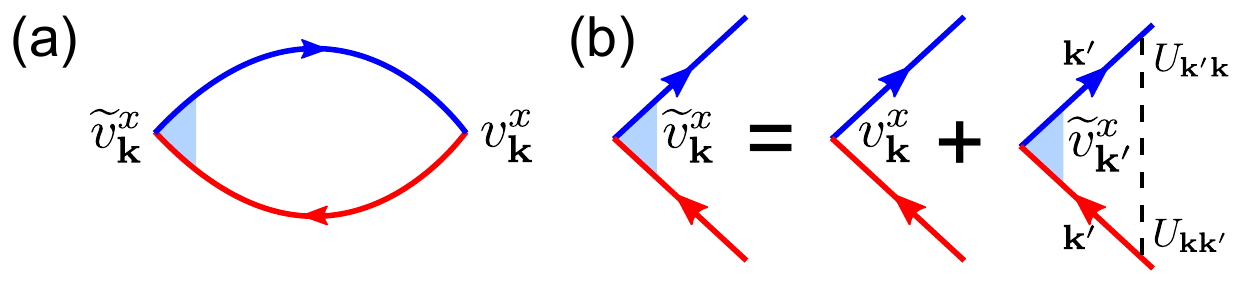}
\caption{(a) The Feynman diagram for the semiclassical (Drude) conductivity $\sigma^{sc}$.
(b) The diagram for the vertex correction to the velocity \cite{Shon98jpsj}. $v$ is the velocity. The arrow lines are for Green's functions. The dashed lines are for disorder scattering ($U$). Replace $x$ by $ z$ for the conductivity along the $z$ direction. }
\label{fig:Drude}
\end{figure}

The total conductivity consists of two parts, the semiclassical Drude conductivity and the quantum interference correction.
In the following sections diagram calculations will be given.
We will first focus on the semiclassical part, which is not sensitive to phase coherences.
With the velocity correction taken into account, the Drude conductivity can be expressed as follows,
\begin{equation}
\sigma_{ii}^{\textrm{sc}}(E_F)=\frac{e^2\hbar }{2\pi }\sum _k\tilde{v}_i v_i G_k^RG_k^A,
\end{equation}
where $\tilde{v}_i$ denote the renormalized velocity due to the impurity scattering,
which are found to be $2 v_z$ and $v_x$ in the $z$ direction and $x$ direction, respectively.
Due to the anisotropy, the Drude conductivities are different in different directions.
\begin{equation}
\sigma _{zz}^{\textrm{sc}}\left(E_F\right)=\frac{e^2\hbar }{2\pi }\sum _k\tilde{v}_zv_zG_k^RG_k^A
=\frac{e^2\hbar }{2\pi }2\sum_kv_zv_zG_k^RG_k^A.
\end{equation}
Converting the summations into integrals and make use of the substitution $v_z=v \cos \theta$, we arrive at
\begin{equation}\label{Drude_z}
\begin{split}
\sigma _{zz}^{\textrm{sc}}\left(E_F\right)&=\frac{e^2\hbar }{\pi }\frac{1}{(2\pi )^3}\int _0^{2\pi }d\varphi \int _0^{\pi }d\theta v^2 \cos^2\theta
\frac{8\pi ^2N_F \tau }{\hbar }=e^2 N_F v^2 \tau.
\end{split}
\end{equation}
Invoking the Einstein relation, $\sigma^{\textrm{sc}}_{zz}=e^2 N_F D_2 $, we have the diffusion constant
$D_2=v^2 \tau=v_z^2 \tau$.
Similarly, $\sigma_{xx}^{\textrm{sc}}$ and $\sigma_{yy}^{\textrm{sc}}$ by symmetry are the same and given by
$\sigma_{xx}^{\textrm{sc}}=e^2N_F D_1$ where $D_1=2 \tau  v_{xf}^2/3 \pi$, $v_{xf}^2=2 E_F/m$, so
$D_1=4 \tau E_F  /3 m\pi= 8 \tau E_F v_{\Vert} /3 \pi$.

The anisotropy thus manifests in different diffusion constants in the in-plane and $z$-direction.

\subsection{Appendix C: Vertex correction to Velocity}\label{velocity_correction}
In this section we will calculate the velocity correction due to impurity scattering. 
For simplicity, we only calculate ``+'' valley; the correction for ``-'' is the same. 
Velocity operators are defined as $\hat{v}_i=(1/\hbar)(\partial H/\partial k_i)$.
In the eigen-basis, they take the following form for ``+'' valley,
\begin{equation}
	\begin{split}
		\bra{+,\bk}\hat{v}_x\ket{+,\bk}=\frac{\sqrt{k \sin^3 \theta} \cos \varphi}{m},\;
		\bra{+,\bk}\hat{v}_y\ket{+,\bk}=\frac{\sqrt{k \sin^3 \theta} \sin \varphi}{m},\;
		\bra{+,\bk}\hat{v}_z\ket{+,\bk}=v \cos \theta.
	\end{split}
\end{equation}
The renormalized velocity can be calculated through the iteration equation,
\begin{equation}\label{velocity_iteration}
\tilde{v}_i=v_i+\sum _{k'} G_{k'}^RG_{k'}^A\langle |U_{k',k}|^2\rangle {}_{\text{imp}}\tilde{v}_{i'},
\end{equation}
where $G^{R/A}$ denote retarded/advanced Green function at zero frequency,
$G^{R/A}(\mathbf{k})=1/(E_F- E_{\mathbf{k}}\pm i \frac{\hbar}{2 \tau}).$
Eq.\eqref{velocity_iteration} can be read off from Fig.\ref{fig:Drude}(b).
The bare diffuson $\langle |U_{k',k}|^2\rangle_{\text{imp}}$ is given by Eq.~(\ref{intra_diff}) and scattering rate $\tau$ is given by Eq.\eqref{intra_rate}.

Assuming the ansatz $\tilde{v}_i=c_i v_i$, 
for $v_z$, we have
\begin{equation}\label{v_z_renormalized}
\begin{split}
c_z\cos \theta _1 &= v \cos \theta _1 +\frac{1}{(2\pi )^3}\int _0^{2\pi }d\varphi _2\int _0^{\pi }d\theta _2\int _0^{\infty
}\frac{k_2}{4m v}dk \frac{1}{E_F-E_{k_2}+\frac{i \hbar}{2\tau }}\frac{1}{E_F-E_{k_2}-\frac{i \hbar}{2\tau }}\langle |U_{k_1,k_2}^{++}|^2\rangle
{}_{\text{imp}}c_z\cos \theta _2\\
&= v \cos \theta _1+c_z\frac{\pi  \tau  \cos \theta _1 N_F n u_0^2}{2 \hbar }=v \cos \theta _1 +\frac{c_z}{2} \cos \theta _1\\
&\Longrightarrow c_z=2,\quad \tilde{v}_z=2 v_z	
\end{split}
\end{equation}
This result is the same as that of 2D Dirac fermions.
For $v_x$ and $v_y$, however, the integrals vanish, i.e., the impurity scattering does not renormalize the velocity in $x$ and $y$ directions.

\subsection{Appendix D: Conductivity correction due to quantum interference}
In this section we focus on the conductivity corrections due to the quantum interference, which can be found as
\begin{equation}
\sigma _{ii}^{\text{qi}}=\sigma _{\text{a1},i}+2\sigma _{\text{a2},i},
\end{equation}
where $\sigma _{\text{a1,2},i}$ stand for the contribution of bare and dressed Hikami boxes and the factor 2 accounts for the fact that there are two equivalent dressed Hikami boxes.
\begin{figure}[ht]
\centering
\includegraphics[width=0.55\textwidth]{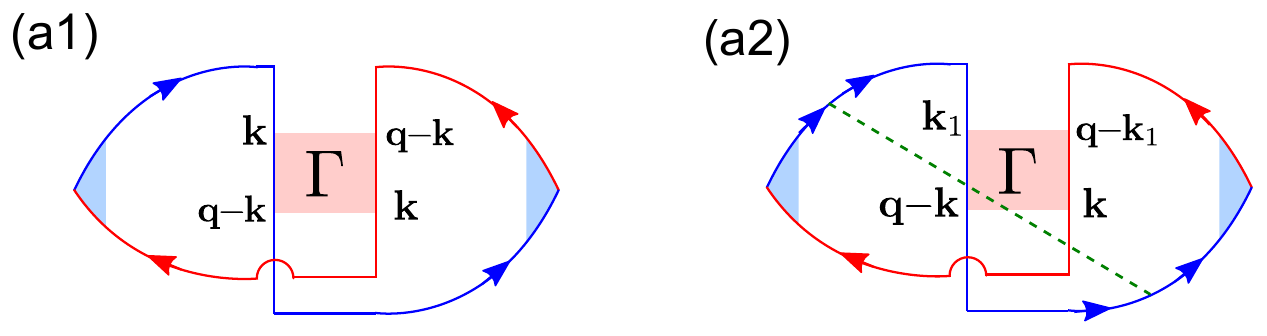}
\caption{The Feynman diagrams for the quantum interference correction to the conductivity that take into account the Cooperons from only the intravalley scattering. These diagrams give $\sigma^{\textrm{qi}}_{ii}$. }
\label{fig:Hikami-boxes-intra}
\end{figure}

Explicitly,
\begin{eqnarray}
\sigma_{a1} &=& \frac{e^2\hbar}{2\pi}\sum_{\mathbf{q}}\Gamma_{\mathbf{k},\mathbf{q}-\mathbf{k}}\sum_{\mathbf{k}}
G^R_{\mathbf{k}} \widetilde{v}^i_\mathbf{k}
G^A_{\mathbf{k}}
G^R_{\mathbf{q-k}}
\widetilde{v}^i_\mathbf{q-k}
G^A_{\mathbf{q-k}}, \nonumber\\
\sigma_{a2}
&=& \frac{e^2  \hbar}{2\pi} \sum_{\mathbf{q}}
\Gamma_{\mathbf{k}_1,\mathbf{q}-\mathbf{k}}
 \sum_{\mathbf{k}}\sum_{\mathbf{k}_1}
\widetilde{v}^i_{\mathbf{k}}\widetilde{v}^i_{\mathbf{q}-\mathbf{k}_1}
G^R_{\mathbf{k}}G^R_{\mathbf{k}_1}G^R_{\mathbf{q}-\mathbf{k}}G^R_{\mathbf{q}-\mathbf{k}_1}G^A_{\mathbf{k}} G^A_{ \mathbf{q}-\mathbf{k}_1}\langle U_{\mathbf{k},\mathbf{k}_1}
U_{\mathbf{q}-\mathbf{k},\mathbf{q}-\mathbf{k}_1}\rangle,
\end{eqnarray}
where $\Gamma_{\mathbf{k},\mathbf{q}-\mathbf{k}}$ and $\langle U_{\mathbf{k},\mathbf{k}_1}
U_{\mathbf{q}-\mathbf{k},\mathbf{q}-\mathbf{k}_1}\rangle$ stand for the full and bare cooperon which are given by Eq.~(\ref{cooperon}) and Eq.~(\ref{bare_cooperon}).
Due to the anisotropy, we need to treat $\sigma_{zz}^{\textrm{qi}}$ and $\sigma_{xx}^{\textrm{qi}}$ separately.

Using Eqs. (\ref{sigma_a1z}) and (\ref{sigma_a2z}), in the $z$ direction, the conductivity correction due to the quantum interference is given by
\begin{equation}\label{qiforz}
\sigma_{zz}^{\text{qi}}=\sigma_{a1,z}+2\sigma_{a2,z}
=-\frac{9 e^2 D_2}{2 \hbar}\sum_{\bq}\frac{1}{ D_1\left(q_x^2+q_y^2\right)+D_2 q_z^2+\omega},
\end{equation}
where
\begin{equation}\label{sigma_a1z}
\begin{split}
\sigma _{\text{a1},z}=-\frac{4 e^2 N_F v^2 \tau ^3}{\hbar ^2}\sum _{\bq}\Gamma _{k,q-k}=-\frac{6 e^2 D_2}{\hbar}\sum_{\bq}\frac{1}{ D_1\left(q_x^2+q_y^2\right)+D_2 q_z^2+\omega},
\end{split}
\end{equation}
and
\begin{equation}\label{sigma_a2z}
\begin{split}
\sigma _{\text{a2},z}
&=\frac{2e^2\hbar }{\pi }\left(-\frac{v^2}{8}\right)\left(-16\pi ^2N_F^2\left(\frac{\tau }{\hbar }\right)^4\right)\frac{\hbar
}{4\pi  N_F \tau } \sum _{\bq} \Gamma_{\bq}\\
&= \frac{3e^2 D_2}{2 \hbar}\sum_{\bq}\frac{1}{ D_1\left(q_x^2+q_y^2\right)+D_2 q_z^2+\omega}.
\end{split}
\end{equation}
Comparing Eqs. \eqref{sigma_a1z} and \eqref{sigma_a2z}, we see the ratio of the dressed one to the bare one $\eta_H= \sigma _{\text{a2}}/\sigma _{\text{a1}}=-1/4$.
On the other hand, the dressed Hikami box for $\sigma_{xx}^{\text{qi}}$ vanishes, hence the quantum interference conductivity correction is found to be
\begin{equation}\label{qiforx}
\begin{split}
\sigma_{xx}^{\text{qi}}=
\sigma _{\text{a1}}=
-\frac{2}{3}e^2 N_F \pi v_{x,F}^2 \frac{\tau^3}{\hbar^2}
\sum_{\bq} \Gamma _{k,q-k}=-\frac{e^2 D_1}{2 \hbar}\sum_{\bq}\frac{1}{ D_1\left(q_x^2+q_y^2\right)+D_2 q_z^2+\omega}.
\end{split}
\end{equation}

The overall minus sign in Eq.~(\ref{qiforz}) and Eq.~(\ref{qiforx}) indicates that, the quantum interference effect reduces the conductivity, i.e., it gives the \emph{weak localization} effect.
In the static limit $\omega=0$, we can evaluate the integral as
\begin{equation}
\begin{split}
&\sum_{\bq}\frac{1}{ D_1\left(q_x^2+q_y^2\right)+D_2 q_z^2}
=\sqrt{\frac{D_1}{D_2}}\int \frac{d \widetilde{q}_z}{2 \pi}\frac{d q_y}{2 \pi}\frac{d q_x}{2 \pi}\frac{1}{D_1 (q_x^2+q_y^2+\widetilde{q}_z^2)}=\frac{1}{2 \pi^2}\sqrt{\frac{D_1}{D_2}}\int_{1/\ell_{\phi}}^{1/\ell_i}\frac{q^2 d q}{D_1 q^2}\\
&=\frac{1}{2 \pi^2}\frac{1}{\sqrt{D_1 D_2}}(\frac{1}{\ell_i}-\frac{1}{\ell_{\phi}}),
\end{split}
\end{equation}
where we have taken the integral bounds to be $(1/\ell_{\phi}, 1/\ell_i)$. Then by using Eq.~(\ref{qiforz}) and Eq.~(\ref{qiforx}) we get
\begin{eqnarray}\label{sigma-qi_zz}
\sigma_{zz}^{\rm {qi}}&=&-\frac{9e^2}{4 \pi^2 \hbar}\sqrt{\frac{D_2}{D_1}}\left(\frac{1}{\ell_z}-\frac{1}{\ell_{\phi}}\right)=-\frac{9e^2}{4 \pi^2 \hbar }\frac{v}{2}\sqrt{\frac{m}{\pi E_F}}\left(\frac{1}{\ell_z}-\frac{1}{\ell_{\phi}}\right),\nonumber\\
\sigma_{xx}^{\rm{qi}}&=&-\frac{e^2}{4 \pi^2 \hbar}\sqrt{\frac{D_1}{D_2}}\left(\frac{1}{\ell_x}-\frac{1}{\ell_{\phi}}\right)=-\frac{e^2}{4 \pi^2 \hbar}\frac{2}{v}\sqrt{\frac{\pi E_F}{m}}\left(\frac{1}{\ell_x}-\frac{1}{\ell_{\phi}}\right).
\end{eqnarray}

\subsection{Appendix E: Magneto-conductivity}
Since the cooperon is anisotropic, the magnetoconductivity also will depend on the orientation of the magnetic field.
For illustration, we consider the case when the field is applied along the $z$-direction, thus $q_x$ and $q_y$ are quantized into Landau levels,
\begin{equation}
q^2_{\Vert}=(n+ \frac{1}{2})\frac{4 e D_1 B}{\hbar}\equiv(n+\frac{1}{2})\frac{1}{\ell_B^2}.
\end{equation}
The upper and lower bounds of the summation can be found as
\begin{equation}
\begin{split}
(n_{\ell_z}+\frac{1}{2})\frac{1}{\ell_B^2} &=\frac{1}{\ell^2_z}\Rightarrow n_l \approx \frac{\ell_B^2}{\ell^2_z}, \quad
(n_{\phi}+ \frac{1}{2})\frac{1}{\ell_B^2}=\frac{1}{\ell_{\phi}^2}\Rightarrow n_{\phi} \approx \frac{\ell_B^2}{\ell^2_{\phi}}.
\end{split}
\end{equation}
The upper cutoff is because $\ell_z$ is the shortest length scale in the diffusive regime. For the lower cutoff, in order to observe the interference effect,
the phase coherence length $\ell_{\phi}\gg \ell_z$ is required.
Then we have
\begin{equation}\label{digamma_function}
\begin{split}
\sum_{\bq}\frac{1}{D_1 q^2_{\Vert}+D_2 q_z^2}&\approx\frac{1}{4 \pi}\int_{-1/\ell_z}^{1/\ell_z}\frac{d q_z}{2 \pi}
\sum^{\ell_B^2/\ell^2_z}_{\ell_B^2/\ell^2_{\phi}}\frac{1}{(n+1/2)+\ell_B^2 q_z^2}\\
&\approx \frac{1}{4 \pi^2}\int_{0}^{1/\ell_z}d q_z \left[\psi\left(\frac{\ell_B^2}{\ell^2_z}+\ell_B^2 q_z^2+\frac{1}{2}\right)-\psi\left(\frac{\ell_B^2}{\ell^2_{\phi}}+\ell_B^2 q_z^2 +\frac{1}{2}\right)\right].
\end{split}
\end{equation}
To reach the second line of Eq.~(\ref{digamma_function}),
we have made use of the property of the digamma function that
\begin{equation}
\psi(x+N)-\psi(x)=\sum_{k=0}^{N-1}\frac{1}{x+k}.
\end{equation}

\subsection{Appendix F: Scattering rate}
In this section we present the derivation of some basic inputs which have been used extensively in this work.
We consider isotropic elastic impurity potentials, which are delta-correlated
\begin{equation}\label{impurity}
\begin{split}
U(\mathbf {r})=\sigma_0 \sum_i u_i \delta(\mathbf{r} -\mathbf{R}_i),\quad \langle U(\mathbf{r})U(\mathbf{r'}\rangle))\propto \delta(\mathbf{r}-\mathbf{r'}).
\end{split}
\end{equation}
Due to the impurity scattering, different eigenstates get mixed.
The general expression for the scattering amplitude can be formulated as
\begin{equation}
\begin{split}
\bra{a,k}U(r)\ket{b,k'}&=\int \bra{a,k}r\rangle\bra{r}U(r)\ket{r'}\bra{r'}b,k'\rangle dr dr'\\
&=\int e^{-i k r}\psi^*(a,k)\sum_i u_i \delta(r'-R_i)\delta(r-r')e^{i k' r'}\psi(b,k)dr dr'\\
&=\sum_i u_i e^{i (k'-k)R_i}\psi^*(a,k)\psi(b,k'),
\end{split}
\end{equation}
where $a,b$ denote the valley degrees of freedom.
Then the correlation of scattering after averaging over impurity configurations can be found as
\begin{equation}
\begin{split}
\langle U_{k_1,k_2}^{ab}U_{k_3,k_4}^{cd}\rangle _{\text{imp}}&=\langle \sum_i u_i e^{i (k_2-k_1)R_i}
\psi^*(a,k_1)\psi(b,k_2) \sum_j u_j e^{i (k_4-k_3)R_j}
\psi^*(c,k_3)\psi(d,k_4)\rangle_{\text{imp}}\\
&= \langle\sum_{i,j}u_i u_j e^{i (k_2-k_1)R_i}e^{i(k_4-k_3)R_j} \rangle_{\text{imp}}\left(\psi^*(a,k_1)\psi(b,k_2)\psi^*(c,k_3)\psi(d,k_4)\right)\\
&\approx n u^2 \delta(k_2-k_1+k_4-k_3)\left(\psi^*(a,k_1)\psi(b,k_2)\psi^*(c,k_3)\psi(d,k_4)\right),
\end{split}
\end{equation}
where the delta function is due to Eq.~(\ref{impurity}); $u$ is assumed to be constant and $n$ is the impurity concentration.
The delta function constraint implies $\bk_2-\bk_1+\bk_4-\bk_3=0$, which can be satisfied by $\bk_2-\bk_1=\bk_3-\bk_4=\bq$, or $\bk_2+\bk_4=\bk_1+\bk_3=\bq$.
The former case is for the diffuson while the latter is for cooperon.

In this work we mainly focus on one valley, i.e., $a=b=c=d$; here we take $a=+$ as an illustration, the results for the $-$ valley are the same,
\begin{equation}\label{intra_diff}
\begin{split}
\langle U_{k_1,k_2}^{++}U_{k_2,k_1}^{++}\rangle _{\text{imp}}&=\frac{n u^2}{4}\left(2+\cos(\theta _1-\theta _2)+\cos(\theta
_1+\theta _2)+2 \cos 2 \left(\varphi _1-\varphi _2\right) \sin \theta _1 \sin \theta _2\right)
\\
&=\frac{\hbar }{16\pi ^4 N_F\tau } \left(2+\cos(\theta _1-\theta _2)+\cos(\theta_1+\theta _2)+2 \cos 2 \left(\varphi _1-\varphi _2\right) \sin \theta _1 \sin \theta _2\right),
\end{split}
\end{equation}
where the second line follows from Eq.~(\ref{intra_rate}).

The calculation of the bare cooperon goes the same way as that of the bare diffuson, except that $\bk_3=-\bk_1$ and $\bk_4=-\bk_2$. In the single valley case it is given by
\begin{equation}\label{bare_cooperon}
\begin{split}
\Gamma _{k_1,k_2}^0&\equiv\langle U_{k_1,k_2}U_{-k_1,-k_2}\rangle _{\text{imp}}
=\frac{n u^2}{4}e^{-2 i \left(\varphi _1-\varphi _2\right)} \left(2+\cos(\theta_1-\theta _2)+\cos(\theta_1+\theta _2)+2 \cos 2 \left(\varphi _1-\varphi _2\right) \sin \theta _1 \sin \theta _2\right)\\
&=\frac{\hbar }{4\pi  N_F\tau} e^{-2 i \left(\varphi _1-\varphi _2\right)} \left(2+\cos(\theta _1-\theta _2)+\cos(\theta_1+\theta _2)+2 \cos 2 \left(\varphi _1-\varphi _2\right) \sin \theta _1 \sin \theta _2\right).
\end{split}
\end{equation}
By using the first-order Born approximation, the intra-valley scattering rate can be calculated as
\begin{equation}\label{intra_rate}
\frac{1}{\tau}=\frac{2\pi }{\hbar }\sum _{k'} \left\langle \left|U_{k',k}|^2\right.\right\rangle {}_{\text{imp}}\delta \left(\omega -\epsilon
_{k'}\right)=\frac{\pi }{ \hbar }n u^2 N_F,
\end{equation}
where $\left\langle \left|U_{k',k}|^2\right.\right\rangle_{\text{imp}}$ is given by Eq.~(\ref{intra_diff}).

\subsection{Appendix G: Cooperon}\label{sec: cooperon}

\begin{figure}[tb]
\centering
\includegraphics[width=0.35\textwidth]{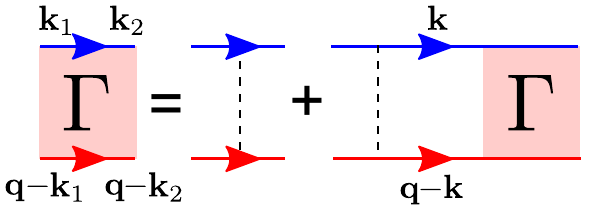}
\caption{The Feynman diagram of the Bethe-Salpeter equation for the intravalley Cooperons. The valley index is suppressed because the valley is conserved during the scattering.}
\label{fig:Cooperon-intra}
\end{figure}

The bare cooperon $\Gamma _{k_1,k_2}^0$ is given by Eq.~(\ref{bare_cooperon}).
The full cooperon can be found by the Bethe-Salpeter equation,
\begin{equation}\label{fullcoop}
\Gamma _{k_1,k_2}=\Gamma _{k_1,k_2}^0+\frac{1}{(2\pi )^3}\int _0^{2\pi }d\varphi \int _0^{\pi }d\theta \int _0^{\infty }\frac{k}{4m v}\Gamma _{k_1,k}^0 G_k^{i\epsilon_n}G_{q-k}^{i \epsilon_n-i \omega_m}\Gamma _{k,k_2}dk,
\end{equation}
where the Matsubara Green function is defined as
\begin{equation}
G\left(k,i \epsilon_n\right)=\frac{1}{i \hbar  \epsilon _n-\xi _k+\frac{i \hbar }{2\tau }\text{sgn}\left(\epsilon _n\right)}.
\end{equation}
Note the bare cooperon is independent of the radial part, so is the ansatz of the full cooperon. We can first deal with the integral
\begin{equation}\label{matsu}
\int _0^{\infty }\frac{k}{4m v}G_k^{i \epsilon_n}G_{q-k}^{i \epsilon_n-i\omega_m}dk=4\pi  N_F\int _{-\infty }^{\infty
}\frac{1}{\xi -i \hbar  \epsilon _n-i\frac{\hbar }{2\tau }}\frac{1}{\xi -i \hbar  \epsilon _n+i \hbar  \omega _m-\hbar \mathbf{v}\cdot \mathbf{q}+i\frac{\hbar }{2\tau }}d\xi.
\end{equation}
In the diffusive limit, Eq.~(\ref{matsu}) can be expanded as
\begin{equation}\label{kernel_expansion}
\int _0^{\infty }\frac{k}{4m v}G_k^{i \epsilon_n}G_{q-k}^{i \epsilon_n-i \omega_m} dk\simeq \frac{8\pi ^2 N_F\tau }{\hbar
 }\left(1-\tau  \omega_m+i \bv \cdot \bq \tau -\left(\bv \cdot \bq\right)^2 \tau ^2 \right).
\end{equation}
The strategy of finding the full cooperon is that, first, make an ansatz of the same structure as the bare one, then iterate Eq.~\eqref{fullcoop} till it closes.
After some straightforward but tedious calculations, the full cooperon turns out to be
\begin{equation}\label{cooperon_in_full}
\begin{split}
\Gamma_{\bk_1,\bk_2}&=\frac{\hbar}{4 N_F\pi \tau} \left(a_1 e^{-2i \left(\varphi _1-\varphi _2\right)}+(a_2-i b_2) e^{-2i \left(\varphi_1-\varphi_2\right)}\sin\theta_1+(a_{11}+i b_{11}) e^{-2i\left(\varphi_1-\varphi_2\right)}\cos\theta_1\right.\\
&\quad+(a_{11}-ib_{11})e^{-2i\left(\varphi_1-\varphi _2\right)}\cos\theta_2+a_{10}e^{-2i\left(\varphi_1-\varphi_2\right)}\cos\theta_1\cos\theta_2+(a_4+ib_4) e^{-4 i \varphi_1+2i\varphi_2}\sin\theta_1\\
&\quad+(a_3-ib_3) e^{2i\varphi_2} \sin\theta_1 +(a_{12}-i b_{12}) e^{-2 i\left(\varphi_1-\varphi_2\right)} \cos\theta_2 \sin\theta_1+(a_{13}+ib_{13}) e^{-4i\varphi_1+2 i \varphi_2} \cos \theta_2 \sin \theta_1\\
&\quad+(a_{14}-ib_{14})e^{2 i\varphi_2} \cos\theta_2 \sin\theta_1+(a_3+i b_3) e^{-2i \varphi_1} \sin \theta_2 +(a_2+i b_2)e^{-2i\left(\varphi_1-\varphi_2\right)} \sin \theta_2 \\
&\quad+(a_4-i b_4) e^{-2 i \varphi_1+4 i \varphi_2} \sin \theta_2+(a_{14}+i b_{14}) e^{-2 i \varphi_1} \cos\theta_1 \sin \theta_2+(a_{13}-i b_{13}) e^{-2 i \left(\varphi _1-2 \varphi _2\right)} \cos \theta_1 \sin\theta_2\\
&\quad+(a_{12}+i b_{12}) e^{-2 i \left(\varphi_1-\varphi_2\right)} \cos \theta_1 \sin\theta_2+a_8 \sin \theta_1  \sin \theta_2 +(a_6-ib_6) e^{-2 i \varphi_1} \sin\theta_1 \sin \theta_2\\
&\quad+(a_9-ib_9) e^{-4 i \varphi_1} \sin \theta_1 \sin \theta_2+(a_5+i b_5) e^{-4i \varphi_1+2 i \varphi_2} \sin \theta_1 \sin \theta_2+(a_5-i b_5) e^{-2 i \varphi_1+4 i \varphi_2} \sin \theta_1 \sin\theta_2\\
&\quad\left.+a_7e^{-4i \varphi_1+4i \varphi_2} \sin \theta_1\sin \theta_2+(a_6+ib_6)e^{2 i \varphi_2} \sin\theta_1\sin\theta_2+(a_9+i b_9) e^{4 i \varphi_2} \sin\theta_1\sin\theta_2\right),	
\end{split}
\end{equation}
where $a_i$ and $b_i$ are real parameters.
In the diffusive limit $\bq \to 0$ and $\omega \to 0$, only $a_1$ term is kept,
\begin{equation}\label{a_1}
a_1\approx \frac{6m^2 \pi}{\tau(2 \tau k_F (q_x^2+q_y^2)+3m^2 \pi(\omega+v^2 \tau q_z^2))},
\end{equation}
correspondingly Eq.~(\ref{cooperon_in_full}) then becomes
\begin{equation}\label{cooperon}
\begin{split}
\Gamma _{\bk_1,\bk_2} &= \Gamma _{\bk,\bq-\bk}\approx \frac{\hbar }{4N_F\pi \tau }e^{- i2\pi}a_1=\frac{\hbar }{4N_F\pi \tau }\frac{6 \pi }{ 2v_{x,F}^2 \tau ^2\left(q_x^2+q_y^2\right)+v^2 q_z^2\tau ^2+3 \pi  \omega  \tau }\\
&\equiv \frac{\hbar }{2 N_F\pi  \tau ^2}\frac{1}{ D_1\left(q_x^2+q_y^2\right)+D_2 q_z^2+\omega }.	
\end{split}
\end{equation}

\end{widetext}

\end{document}